\pdfoutput=1
\documentclass[%
superscriptaddress,
nofootinbib,
twocolumn,
amsmath,amssymb,
aps,
prd
]{revtex4-2}

\usepackage{graphicx}
\usepackage{dcolumn}
\usepackage{bm}

\def\apj{{ApJ}}                 
\def\aap{{A\&A}}                
\def\baas{{BAAS}}               

\def\jcap{{J. Cosmology Astropart. Phys.}}
\def\mnras{Mon.~Not.~Roy.~Astron.~Soc.}             
\def\prd{{Phys.~Rev.~D}}        
\def\physrep{{Phys.~Rep.}}   

\newcommand{\beq}{\begin{equation}}
\newcommand{\beqa}{\begin{eqnarray}}
\newcommand{\eeq}{\end{equation}}
\newcommand{\eeqa}{\end{eqnarray}}

\newcommand{\simgt}{\lower.5ex\hbox{$\; \buildrel > \over \sim \;$}}
\newcommand{\simlt}{\lower.5ex\hbox{$\; \buildrel < \over \sim \;$}}

\newcommand{\bd}[1]{\mbox{\boldmath $#1$}}
\newcommand{\tj}[6]{ 
\begin{pmatrix}
   #1 & #2 & #3 \\
   #4 & #5 & #6 
\end{pmatrix}
}
  
\usepackage[usenames,dvipsnames]{xcolor}
\newcommand{\rev}[1]{{\color{black} #1}}

\begin{document}


\title{Constraining Primordial Non-Gaussianity with \\
Post-reconstructed Galaxy Bispectrum in Redshift Space}

\author{Masato Shirasaki}
 \email{masato.shirasaki@nao.ac.jp}
\affiliation{%
National Astronomical Observatory of Japan (NAOJ), Mitaka, Tokyo 181-8588, Japan
}%
\affiliation{
The Institute of Statistical Mathematics,
10-3 Midori-cho, Tachikawa, Tokyo 190-8562, Japan
}
\author{Naonori S. Sugiyama}%
 \email{naonori.sugiyama@nao.ac.jp}
\affiliation{%
National Astronomical Observatory of Japan (NAOJ), Mitaka, Tokyo 181-8588, Japan
}%


\author{Ryuichi Takahashi}
\affiliation{
Faculty of Science and Technology, Hirosaki University, 3 Bunkyo-cho, Hirosaki, Aomori, 036-8561, Japan
}%

\author{Francisco-Shu Kitaura}
\affiliation{%
Instituto de Astrof\'isica de Canarias, s/n, E-38205, La Laguna, Tenerife, Spain
}%
\affiliation{
Departamento de Astrof\'isica Universidad de La Laguna, E-38206, La Laguna, Tenerife, Spain
}


\date{\today}

\begin{abstract}
Galaxy bispectrum is a promising probe of 
inflationary physics in the early universe 
as a measure of primordial non-Gaussianity (PNG), 
whereas its signal-to-noise ratio 
is significantly affected 
by the mode coupling due to non-linear gravitational growth.
In this paper, we examine 
the standard reconstruction method 
of linear cosmic mass density fields 
from non-linear galaxy density fields to 
de-correlate the covariance in redshift-space galaxy bispectra. 
In particular, we evaluate the covariance of the bispectrum for massive-galaxy-sized dark matter halos 
with reconstruction by using 4000 independent $N$-body simulations.
Our results show that 
the bispectrum covariance for the post-reconstructed field approaches the Gaussian prediction 
at scale of 
$k<0.2\, h\, {\rm Mpc}^{-1}$.
We also verify
the leading-order PNG-induced bispectrum
is not affected by details of the reconstruction with 
perturbative theory.
We then demonstrate the constraining power of the post-reconstructed bispectrum for PNG 
at redshift of $\sim0.5$. 
Further, we perform a Fisher analysis to make a forecast of PNG constraints by galaxy bispectra including anisotropic signals. 
Assuming a massive galaxy sample in the 
SDSS Baryon Oscillation Spectroscopic Survey, 
we find that the post-reconstructed bispectrum 
can constrain the local-, 
equilateral-
and orthogonal-types of PNG with 
$\Delta f_{\rm NL} \sim$13, 90 and 42, respectively, 
improving the constraints with the pre-reconstructed bispectrum by a factor of $1.3-3.2$.
In conclusion, the reconstruction plays an essential role in constraining various types of 
PNG signatures
with a level of 
$\Delta f_{\rm NL}\simlt 1$ 
from the galaxy bispectrum
based on
upcoming galaxy surveys.
\end{abstract}

\maketitle


\section{Introduction}
The origin of large-scale structures in the universe is a key question in modern cosmology.
Inflation is among the best candidates for the production mechanism of the seeds 
of primordial density fluctuations, while the physics behind 
inflation is still unclear.
A deviation from Gaussianity in 
the initial curvature perturbations, referred to as primordial non-Gaussianity (PNG), is considered to be a unique quantity to constrain the physics of 
inflationary models in the early universe \cite{Bartolo:2004if}. 
The degree of PNG is observable 
by measuring 
the three-point correlation function in the anisotropy of 
the cosmic microwave 
background (CMB) \cite{Komatsu:2003iq}, as well as in the spatial distribution of tracers of large-scale structures, e.g., galaxies \cite{Sefusatti:2007ih}.

The tightest constraint on PNG has been obtained 
from
the statistical analysis of CMB measured by 
the Planck satellite \cite{Ade:2015ava}. 
Future galaxy surveys have great potential in
improving the Planck constraints 
by a factor of $\sim10$ \cite{Karagiannis:2018jdt, 2019BAAS...51c..72F}.
In particular, the three-point correlation analysis of galaxies will be 
key for the next breakthrough in our understanding of the early universe, because it enables us to explore a wider range of inflationary models than the conventional two-point correlations.
\rev{Most previous studies on PNG forecasts assume that the covariance of galaxy bispectra (i.e.~the three-point correlation in Fourier space) follows Gaussian statistics
(e.g.~Ref.~\cite{2019BAAS...51c..72F} and see references therein)\footnote{Recently, Ref.~\cite{2020JCAP...06..041G} presented forecasts of standard cosmological parameter constraints with redshift-space galaxy power spectra and bispectra by including non-Gaussian covariances.
Nevertheless, the impact of the non-Gaussian covariances on the PNG constraint is still unclear.}, 
whereas it is not always valid in real.}
Non-linear gravitational growth can naturally induce additional correlated scatters
among different length scales in measurements of galaxy bispectra, referred to as 
non-Gaussian covariances.
Surprisingly, the non-Gaussian covariance can dominate the statistical error of the galaxy bispectrum 
even at length scales of $\sim 100\, {\rm Mpc}$, leading to the degradation of 
the signal-to-noise ratio
by a factor of $3-4$ in the current galaxy surveys \cite{Chan:2016ehg, 2020MNRAS.497.1684S}.
Hence, the actual constraining power of the galaxy bispectrum for PNG will hinge on
details in sample covariance estimation.

In this paper, 
we investigate the possibility of reducing the non-Gaussian covariance and 
increasing the information 
content in the galaxy bispectrum
by using a reconstruction method as developed in Ref~\cite{Eisenstein:2006nk}. 
The original motivation of this method was 
to obtain precise measurements of the baryon acoustic oscillations (BAO) by 
\rev{reducing non-linear gravitational effects in observed galaxy density fields}, 
effectively linearizing the two-point statistics.
\rev{This reconstruction method has been applied to the measurement of three-point correlations in the Sloan Digital Sky Survey (SDSS), allowing to enhance the acoustic feature in the non-Gaussian observable at large scales \cite{2017MNRAS.469.1738S}.}
Apart from that initial benefit, 
we show that reconstruction can reduce
the correlated scatters in the observed galaxy bispectra, as it removes  
non-linear mode coupling in the galaxy density field on large scales \cite{Padmanabhan:2008dd}. 
Recently, Ref.~\cite{2020arXiv200713998H} found the reduction of the correlated scatters in the power spectrum for post-reconstructed cosmic mass density fields, whereas
we extend the previous study to halo density fields.
We evaluate the covariance of post-reconstructed bispectrum among 4000 $N$-body simulations and demonstrate how much gain in the information contents in the post-reconstructed bispectrum will be obtained without adding new survey volumes.

Apart from the covariance, we also study 
the information content
in anisotropic components of galaxy bispectra caused 
by redshift-space distortions (RSDs).
To increase the signal-to-noise ratio of PNG in galaxy surveys, 
we need precise constraints on the galaxy bias.
Because the information of non-linear velocity fields is imprinted in the anisotropic bispectrum and 
is independent of non-linear 
bias, we expect that the anistropic bispectrum will solve some degeneracies between PNG and galaxy-bias parameters.
For this purpose, we adopt a framework in Ref.~\cite{Sugiyama:2018yzo} to decompose
the galaxy bispectra into isotropic and anisotropic components 

The rest of the present paper is organized as follows.
In Section~\ref{sec:sim}, we describe our simulation data to study covariance matrices of the post-reconstructed bispectrum
and how to measure the bispectrum from the simulation.
In Section~\ref{sec:model}, we summarize a theoretical model to predict statistics of post-reconstructed density fields.
We present the results in Section~\ref{sec:results}.
Concluding remarks and discussions are given in Section~\ref{sec:con}.

\section{Simulation and Method}\label{sec:sim}

To study the post-reconstructed galaxy bispectrum, 
we run 4000 independent realizations of a cosmological $N$-body simulation. 
We perform the simulation with  
{\tt Gadget-2} Tree-Particle Mesh code \cite{Springel:2005mi}.
Each simulation contains $512^3$ particles in a cubic volume of $500^3\, (h^{-1}\, {\rm Mpc})^3$.
We generate the initial conditions using a parallel code developed by Refs.~\cite{2009PASJ...61..321N, 2011A&A...527A..87V}, which employs the second-order Lagrangian perturbation theory \cite{Crocce:2006ve}.
We assume that the initial density fluctuations follow Gaussian statistics.
The initial redshift is set to $z_{\rm init}=31$, 
where we compute the linear matter transfer function using {\tt CAMB} \citep{Lewis:1999bs}. 
We adopt the following parameters in the simulations: 
present-day matter density parameter $\Omega_{\rm m0} = 0.3156$, 
dark energy density $\Omega_{\Lambda} = 1-\Omega_{\rm m0} = 0.6844$, 
the density fluctuation amplitude $\sigma_{8} = 0.831$, 
the parameter of the equation of state of dark energy $w_{0}=-1$, 
Hubble parameter $h=0.6727$, 
and the scalar spectral index $n_s=0.9645$. 
These parameters are consistent with the Planck 2015 results \cite{Ade:2015xua}.
We output the simulation data at $z=0.484$, 
which represents the intermediate redshift for available luminous red galaxy catalogs
from the SDSS Baryon Oscillation Spectroscopic Survey (BOSS)
\cite{Dawson:2012va}.

We then identify the dark matter halos from 
the corresponding simulations
using the phase-space temporal halo finder {\tt ROCKSTAR} \cite{2013ApJ...762..109B}.
In the following, we consider a sample of dark matter halos with a mass range of $10^{13-14}\, h^{-1}\, M_{\odot}$ at $z=0.484$ \footnote{We define the halo mass by spherical over-density mass with respect to
200 times mean matter density in the universe.}. 
Note that the mass range of our halo sample is similar to one in a galaxy sample in BOSS \cite{Miyatake:2013bha}.
The average number density of this halo sample over 4000 realizations
is found to be $3.9\times10^{-4}\, (h^{-1}\, {\rm Mpc})^{-3}$. 
Individual halos in our sample are resolved by $126-1260$ particles.
To take into account 
the effect of 
redshift space distortions caused by the peculiar velocity field in the clustering analysis,
we set the $z$-axis in our simulation to be the line-of-sight direction and work with the distant-observer approximation.

Throughout this paper, we follow the ``standard" reconstruction method as in Ref~\cite{Eisenstein:2006nk}.
When computing the displacement field, we apply a Gaussian filter with a smoothing scale 
of $R=10\, h^{-1}\, {\rm Mpc}$ and divide the resulting smoothed density field by the linear bias 
$b_{\rm fid}=1.8$, according to our halo sample \cite{2010ApJ...724..878T}. 
To be specific, the displacement vector is defined in Fourier space as
\begin{equation}
    \bd{s}(\bd{k}) = -i \frac{1}{b_{\rm fid}}\frac{\bd{k}}{k^2} W(k,R) \delta_{\rm h}(\bd{k}), \label{eq:s_ZA}
\end{equation}
where $\delta_{\rm h}$ is the halo density fluctuation including RSDs, 
$W(k,R)=e^{-k^2R^2/2}$ is the Gaussian smoothing function.
Note that we do not remove linear RSDs through reconstruction.
We grid halos onto meshes with $512^3$ cells using the cloud-in-cell assignment scheme.
Furthermore, in our statistical analyses,
we use the randoms 100 times as many point sources as halos for a given realization in our simulation volume.
\rev{Under the Zel'dovich approximation, the displacement field of Eq.~(\ref{eq:s_ZA}) sets the density field to be in its initial state.
Hence, the reversal Zel'dovich method will provide a means of reducing the non-linear gravitational growth in the density field of interest.
We study this effect in details in Section~\ref{sec:model}.}

To compute the redshift-space bispectrum from the simulation data, we follow a decomposition formalism developed in Ref.~\cite{Sugiyama:2018yzo}.
This approach is efficient to separate 
the anisotropic and isotropic signals 
from the observed bispectrum.
For a given halo overdensity field, one defines the halo bispectrum as 
\beqa
\langle \delta_{\rm h}(\bd{k}_1)
\delta_{\rm h}(\bd{k}_2) 
\delta_{\rm h}(\bd{k}_3)
\rangle
&\equiv& (2\pi)^3 
\delta_{\mathrm{D}}(\bd{k}_1+\bd{k}_2+\bd{k}_3) \nonumber \\
&&
\quad \quad \quad \quad
\times B(\bd{k}_1, \bd{k}_2, \bd{k}_3),
\eeqa
where $\delta_{\mathrm{D}}$ 
represents the Dirac delta function,
and $B(\bd{k}_1, \bd{k}_2, \bd{k}_3)$ is the bispectrum.
In redshift space, the bispectrum depends on the line-of-sight direction of each halo as well, causing anisotropic signals.
Ref.~\cite{Sugiyama:2018yzo} found that such anisotropic signals in the redshift-space bispectrum can be well characterized with a tri-polar spherical harmonic basis \cite{1988qtam.book.....V}.
The coefficient in the tri-polar spherical harmonic decomposition of the redshift-space bispectrum is given by
\beqa
B_{\ell_1\, \ell_2\, L}(k_1, k_2)
&=& 
\int \frac{\mathrm{d}\hat{k}_1}{4\pi}
\int \frac{\mathrm{d}\hat{k}_2}{4\pi}
\int \frac{\mathrm{d}\hat{n}}{4\pi} 
{\cal W}_{\ell_1\, \ell_2\, L}(\hat{k}_1, \hat{k}_2, \hat{n})
\nonumber \\
&&
\quad \quad \quad \quad \quad 
\times B(\bd{k}_1, \bd{k}_2, -\bd{k}_{12}), \\
{\cal W}_{\ell_1\, \ell_2\, L}(\hat{k}_1, \hat{k}_2, \hat{n})
&\equiv& 
(2\ell_1+1)(2\ell_2+1)(2L+1)
\tj{\ell_1}{\ell_2}{L}{0}{0}{0} \nonumber \\
&& 
\times
\sum_{m_1, m_2, M}\, 
\tj{\ell_1}{\ell_2}{L}{m_1}{m_2}{M}
\nonumber \\
&& 
\quad \quad \quad \quad
\times
y^{m_1}_{\ell_1}(\hat{k}_1)\, 
y^{m_2}_{\ell_2}(\hat{k}_2)\, 
y^{M}_{L}(\hat{n}),
\eeqa
where 
$\bd{k}_{12} = \bd{k}_1+\bd{k}_2$,
$y^{m}_{\ell} = \sqrt{4\pi/(2\ell+1)}\, Y^{m}_{\ell}$
is a normalized spherical harmonic function,
and $\left(\cdots \right)$ with six indices represents the Wigner-3$j$ symbol.
In this decomposition, the index $L$ governs the expansion with respect to the line-of-sight direction.
The mode of $B_{\ell_1\, \ell_2\, L}$ with $L=0$ describes
isotropic components in the bispectrum, 
while the modes with $L>0$ arise from anisotropic components alone.
Hence, we refer $B_{\ell_1\, \ell_2\, L}$ with $L=0$ and $L=2$ 
to as the monopole and quadrupole bispectra, respectively.

In this paper, we consider 
the lowest-order monopole bispectrum 
$B_{000}(k_1, k_2)$ 
and an anisotropic term $B_{202}(k_1, k_2)$.
It would be worth noting that $B_{202}$ is a leading anisotropic signal 
for a sample of massive galaxies \cite{Sugiyama:2018yzo}.
When measuring the bispectrum, we employ 
the linear binning in $k_{i}$ ($i=1,2$) 
for the range of $0.01-0.3\, h\, {\rm Mpc}^{-1}$ with 
the number of bins being 15.
Hence, the total number of degrees of freedom in $B_{000}$ is $15\times(15+1)/2 = 120$,
while $B_{202}$ consists of $15\times15=225$ data points.
We apply the three-dimensional Fast Fourier Transform (FFT) 
on $512^3$ grids
by using the triangular-shaped cloud assignment.
The details of the algorithm for bispectrum 
measurements are found in Ref.~\cite{Sugiyama:2018yzo}.
We evaluate the covariance using 4000 realizations of our halo samples with reconstruction, as well as in the absence of reconstruction\footnote{When inverting the covariance, we take into account the correction as in Ref.~\cite{Hartlap:2006kj}. The correction is found to be of an order of $8\%$ at most 
in our analysis.}.

\section{Model}\label{sec:model}

\subsection{Perturbative approach}

We develop an analytic model to predict the statistical properties of the reconstructed density field.
\rev{In the standard reconstruction method, we shift halo number density fields as well as random data points with a given displacement field \bd{s} \cite{Eisenstein:2006nk}. This process is formally written as
\beqa
n^{(\rm rec)}_{\rm h}(\bd{x}) &=& \int {\rm d}^3 x^{\prime}
n_{h}(\bd{x}^{\prime}) \,
\delta_{\rm D}(\bd{x}-\bd{x}^{\prime}-\bd{s}(\bd{x}^{\prime})), \label{eq:n_h_rec} \\
n^{(\rm rec)}_{\rm r}(\bd{x}) &=& \int {\rm d}^3 x^{\prime}
\bar{n}_{h}\,
\delta_{\rm D}(\bd{x}-\bd{x}^{\prime}-\bd{s}(\bd{x}^{\prime})),
\label{eq:n_r_rec}
\eeqa
where 
$\delta_{D}$ is the 3D Dirac delta function, 
$n_{\rm h}$ is the halo number density of interest, $\bar{n}_{\rm h}$ is the mean halo number density,
$n^{(\rm rec)}_{\rm h}$ and $n^{(\rm rec)}_{\rm r}$ are the post-reconstructed halo number density and random fields, respectively.
}
\rev{Using Eqs.~(\ref{eq:n_h_rec}) and (\ref{eq:n_r_rec}), we can relate the reconstructed density field $\delta^{(\rm rec)}_{\rm h}$ with the pre-reconstructed counterpart $\delta_{\rm h}$ as 
\beqa
\delta^{(\rm rec)}_{\rm h}(\bd{x}) &=& 
\frac{n^{(\rm rec)}_{\rm h}(\bd{x})-n^{(\rm rec)}_{\rm r}(\bd{x})}{\bar{n}_{\rm h}} \nonumber \\
&=&
\int {\rm d}^3 x^{\prime} \delta_{\rm h}(\bd{x}^\prime) \, 
\delta_{\rm D}(\bd{x}-\bd{x}^{\prime}-\bd{s}(\bd{x}^{\prime})),
\eeqa
where $\delta_{\rm h} = n_{\rm h}/\bar{n}_{\rm h} -1$.
}
The Fourier counterpart of $\delta^{(\rm rec)}_{\rm h}$ is then given by
\beqa
\delta^{(\rm rec)}_{\rm h}(\bd{k}) 
&=& \int \mathrm{d}^3 x\, e^{-i\bd{k}\cdot\bd{x}}
e^{-i \bd{k}\cdot \bd{s}(\bd{x})} \delta_{\rm h}(\bd{x}) \nonumber \\ 
&=& 
\delta_{\rm h}(\bd{k})
+
\sum_{n=1}^{\infty}\,
\frac{(-1)^{n}}{n!}
\int\frac{\mathrm{d}^3 k_{1}}{(2\pi)^3}
\cdots
\int\frac{\mathrm{d}^3 k_{n}}{(2\pi)^3} \nonumber \\
&&
\quad \quad \quad
\times\,
\left[\bd{k}\cdot\bd{{\cal S}}(\bd{k}_1)\right]
\cdots
\left[\bd{k}\cdot\bd{{\cal S}}(\bd{k}_n)\right] \nonumber \\
&&
\quad \quad \quad
\times\,
\delta_{\rm h}(\bd{k}-\bd{k}_{1 \cdots n})
\delta_{\rm h}(\bd{k}_1)
\cdots
\delta_{\rm h}(\bd{k}_n), \label{eq:delta_h_rec_f}
\eeqa
where $\bd{k}_{1\cdots n} = \bd{k}_1+\cdots+\bd{k}_{n}$
and we introduce 
$\bd{{\cal S}}(\bd{k})=\bd{k}\, W(k)/(b_{\rm fid} k^2)$.
In redshift space, 
the standard perturbation theory predicts
\cite{Scoccimarro:1999ed}
\beqa
\delta_{\rm h}(\bd{k}) &=& \sum_{n=1}^{\infty}
\int \frac{\mathrm{d}^3 k_{1}}{(2\pi)^3} \cdots 
\int \frac{\mathrm{d}^3 k_{n}}{(2\pi)^3}
\, \delta_{\mathrm{D}}(\bd{k}-\bd{k}_{1\cdots n}) \nonumber \\
&&
\quad \quad
\times\, 
Z_{n}(\bd{k}_{1}, \cdots, \bd{k}_{n})
\, 
\delta_{\rm L}(\bd{k}_{1})
\cdots
\delta_{\rm L}(\bd{k}_{n}), \label{eq:delta_h_ST}
\eeqa
where 
$Z_{n}(\bd{k}_{1},\cdots,\bd{k}_{2})$ represents the $n$-th order kernel function (the details are found in Ref.~\cite{Scoccimarro:1999ed}),
and $\delta_{\rm L}$ is the linear mass density perturbation.
Using Eqs.~(\ref{eq:delta_h_rec_f}) and (\ref{eq:delta_h_ST}),
we find that the post-reconstructed halo over-density field is expressed in a similar manner to Eq.~(\ref{eq:delta_h_ST}), 
\beqa
\delta^{(\mathrm{rec})}_{\rm h}(\bd{k}) &=& \sum_{n=1}^{\infty}
\int \frac{\mathrm{d}^3 k_{1}}{(2\pi)^3} \cdots 
\int \frac{\mathrm{d}^3 k_{n}}{(2\pi)^3}
\, \delta_{\mathrm{D}}(\bd{k}-\bd{k}_{1\cdots n}) \nonumber \\
&&
\times\, 
Z^{(\mathrm{rec})}_{n}(\bd{k}_{1}, \cdots, \bd{k}_{n})
\, 
\delta_{\rm L}(\bd{k}_{1})
\cdots
\delta_{\rm L}(\bd{k}_{n}), \label{eq:delta_h_rec_ST}
\eeqa
where 
\beqa
Z^{(\mathrm{rec})}_{n}
&\equiv& Z_{n}+ \Delta Z_{n}, \\
\Delta Z_{1}(\bd{k}) &=& 0, \label{eq:Zrec_1} \\
\Delta Z_{2}(\bd{k}_1, \bd{k}_2)
&=& 
-\frac{1}{2} 
\left[
\bd{k}_{12} \cdot \bd{{\cal S}}(\bd{k}_1) 
+ 
\bd{k}_{12} \cdot \bd{{\cal S}}(\bd{k}_2)
\right] \nonumber \\
&&
\quad \quad \quad \quad
\times
Z_{1}(\bd{k}_1) Z_{1}(\bd{k}_2), \label{eq:Zrec_2} \\
\Delta Z_{3}(\bd{k}_1,\bd{k}_2,\bd{k}_3)
&=& 
-\frac{1}{3} 
\left[
\bd{k}_{123}\cdot\bd{{\cal S}}(\bd{k}_{12}) 
+ 
\bd{k}_{123}\cdot\bd{{\cal S}}(\bd{k}_{3})
\right] \nonumber \\
&&
\quad \quad \quad \quad
\times
Z_{2}(\bd{k}_1,\bd{k}_2)\, Z_{1}(\bd{k}_3) \nonumber \\
&&
\quad
+
\frac{1}{6}
\left[
\left(\bd{k}_{123}\cdot\bd{{\cal S}}(\bd{k}_1)\right)
\left(\bd{k}_{123}\cdot\bd{{\cal S}}(\bd{k}_2)\right)
\right]\nonumber \\
&&
\quad \quad \quad \quad
\times
Z_{1}(\bd{k}_1)Z_{1}(\bd{k}_2)Z_{1}(\bd{k}_3) 
\nonumber\\
&&
\quad
+\mbox{(2 cyc.)}, \label{eq:Zrec_3}
\eeqa
and so on.

Following Ref.~\cite{2009JCAP...08..020M},
we expand the (pre-reconstructed) halo over-density field into the underlying mass density fluctuation $\delta_{\rm m}$ up to the second order\footnote{
Note that $\delta_{\rm m}$ can be expanded in perturbation theory as well. It holds that $\delta_{\rm m} = \delta_{\mathrm{L}}$ at the lowest order in the standard perturbation theory \cite{2002PhR...367....1B}.}
\beqa
\delta_{\rm h}(\bd{k}) 
&=& 
b_{1} \, \delta_{\rm m}(\bd{k})
+ \int\, \frac{\mathrm{d}^3q}{(2\pi)^3}\,
\delta_{\rm m}(\bd{q})\, \delta_{\rm m}(\bd{k}-\bd{q}) \nonumber \\
&&
\quad \quad \quad \quad 
\times\left[\frac{b_2}{2} + b_{K_2}\, K(\bd{q}, \bd{k}-\bd{q})\right], \label{eq:delta_h_exp}
\eeqa
where  
$b_1$ stands for the linear bias, $b_2$ represents the second-order local bias, $b_{K_2}$ is the tidal bias, 
and the function $K$ is given by
\beqa
K(\bd{k}_1, \bd{k}_2) = \frac{\bd{k}_1 \cdot \bd{k}_2}{k_1\, k_2} - \frac{1}{3}.
\eeqa
In this notation of the galaxy biasing, 
$Z_{1}(\bd{k}) = b_1 + f \mu^2$ corresponds to the Kaiser formula of linear RSDs \cite{Kaiser:1987qv}
with $f$ and $\mu$ being the logarithmic growth rate function and the cosine between wavevector and the line-of-sight direction.
Note that $Z_{2}$ includes non-linear bias terms such as
$b_2/2 + b_{K_2}K(\bd{k}_1, \bd{k}_2)$.
We find that 
the leading-order fluctuation in 
the post-reconstructed field is
independent of the detail in the assumed displacement field
(e.g. a choice of the filter function 
$W$ in Eq.~[\ref{eq:s_ZA}]).
For the case of dark matter in real space, 
we can reproduce previously known forms in Ref.~\cite{Schmittfull:2015mja,Hikage:2017tmm}
by replacing the kernel functions $Z_{n}$ with
\beqa
Z_{1} \to 1 \, \mathrm{and} \, Z_{n\geq2} \to F_{n\geq2},
\eeqa
where the function $F_{n}$ represents the $n$-th order kernel function in the standard perturbation theory for the dark matter density field (see Ref.~\cite{2002PhR...367....1B} for a review).

\subsection{Bispectrum}

We then consider the bispectrum of the post-reconstructed field up to corrections ${\cal O}(\delta_{\rm L}^5)$ as
\begin{eqnarray}
B (\bd{k}_1,\bd{k}_2,\bd{k}_3)
= B_{3} + B_{4} + B_{5}
\end{eqnarray}
where $B_{n}$ is the bispectrum 
with an order of ${\cal O}(\delta_{\rm L}^{n})$.
Note $B_{n={\rm odd}}$ arises from PNG, because they become zero if the density fluctuation was purely Gaussian,
and $B_{4}$ is generated by both the non-linear gravity and 
a primordial 4-point function in general.
In this paper, we ignore the term coming from the primordial 4-point function for simplicity.

The linear matter perturbations at a redshift $z$ , denoted as $\delta_{\rm L}(\bd{k}, z)$, 
is related to the curvature perturbation $\Phi(\bd{k})$ 
through the function $M(k, z)$ as 
\beqa
\delta_{\rm L}(\bd{k}, z) &=& M(k, z)\, \Phi(\bd{k}), \\
M(k,z) &=& \frac{2k^2 c^2 T(k) D(z)}{3\Omega_{\rm m0}H^2_0},
\eeqa
where $D(z)$ is the linear growth factor\footnote{We normalize $D$ to unity today, i.e. $D(z=0) = 1$},
$T(k)$ is the matter transfer function normalized to unity 
on large scales $k\rightarrow0$ and $c$ is the speed of light. Note that we omit the redshift $z$ in the following for simplicity.
The leading term $B_{3}$ is given by
\begin{eqnarray}
	B_{3}(\bd{k}_1,\bd{k}_2,\bd{k}_3)
	&=& 
	Z_{1}({\bd{k}_1)Z_{1}(\bd{k}_2})Z_{1}(\bd{k}_3) 
	M(k_1)\, M(k_2)\, M(k_3) \nonumber \\
	&&
	\quad \quad \quad \quad
	\quad \quad \quad \quad
	\times
	B_{\Phi}(\bd{k}_1,\bd{k}_2,\bd{k}_3), \label{eq:B3}
\end{eqnarray}
where $B_{\Phi}$ represents 
the bispectrum of the primordial curvature perturbation.
Note that Eq.~(\ref{eq:B3}) is valid even for the post-reconstructed field, because
it holds $Z_{1} = Z^{(\mathrm{rec})}_{1}$.

The first gravity-induced term $B_{4}$ 
is well-known as the tree-level solution.
For the post-reconstructed field, it is given by 
\begin{eqnarray}
	B_{4}(\bd{k}_1,\bd{k}_2,\bd{k}_3) 
	&=& 
	Z_{1}(\bd{k}_1)\, Z_{1}(\bd{k}_2)\,
	Z^{(\mathrm{rec})}_{2}(\bd{k}_1,\bd{k}_2) \nonumber \\
	&&
	\quad 
	\times P_{\rm L}(k_1)\, P_{\rm L}(k_2)
	+ \mbox{(2 cyc.)}, \label{eq:B4}
\end{eqnarray}
where $P_{\rm L}(k) = M^2(k) P_{\Phi}(k)$ with $P_{\Phi}$ being the curvature perturbation power spectrum.
For the pre-reconstructed field, we can obtain the solution by replacing $Z^{(\mathrm{rec})}_{2}$ with $Z_{2}$ as well.
We finally express the term $B_{5}$, 
referred to as the 1-loop correction, as 
\begin{eqnarray}
	&&
	B_{5}(\bd{k}_1,\bd{k}_2,\bd{k}_3) \nonumber \\
	&&
	=
	4\int_{\bd{p}} 
	  Z^{(\rm rec)}_{2}(\bd{k}_1-\bd{p},\bd{p})
	  Z^{(\rm rec)}_{2}(\bd{k}_2+\bd{p},-\bd{p})
	  Z_{1}(\bd{k}_3) \nonumber \\
	&&
	\quad \quad
	\times\, 
	  P_{\rm L}(p) 
	  B_{\rm pri}(\bd{k}_1-\bd{p}, \bd{k}_2+\bd{p}, \bd{k}_3)
	  + \mbox{\small{(2 cyc.)}}\nonumber \\
	&&
	+\,
	2 \int_{\bd{p}} 
	  Z^{(\rm rec)}_{2}(\bd{k}_1,\bd{k}_3)
	  Z^{(\rm rec)}_{2}(\bd{k}_1-\bd{p},\bd{p})
	  Z_{1}(\bd{k}_3) \nonumber \\
	&&
	\quad \quad
	\times\, 	  
	  P_{\rm L}(k_3) 
	  B_{\rm pri}(\bd{k}_1-\bd{p}, -\bd{k}_1, \bd{p}) 
	  + \mbox{\small{(5 cyc.)}}\nonumber \\
	&& 
	+\, 
	3 \int_{\bd{p}} 
	Z_{1}(\bd{k}_1) Z_{1}(\bd{k}_2)
	Z^{({\rm rec})}_{3}(\bd{k}_2+\bd{p}, \bd{k}_1,-\bd{p}) 
	\nonumber \\
	&&
	\quad \quad
	\times\, 
	P_{\rm L}(k_1) 
	B_{\rm pri}(\bd{k}+\bd{p}, -\bd{k}_2,\bd{p}) 
	+ \mbox{\small{(5 cyc.)}} 
	\nonumber \\
	&&
	+\, 
	3 \int_{\bd{p}} 
	Z_{1}(\bd{k}_1) Z_{1}(\bd{k}_2)
	Z^{({\rm rec})}_{3}(\bd{k}_3, \bd{p}, -\bd{p}) 
	\nonumber \\
	&&
	\quad \quad
	\times\, 
	P_{\rm L}(p) 
	B_{\rm pri}(\bd{k}_1,\bd{k}_2,\bd{k}_3) 
	+ \mbox{\small{(2 cyc.)}}, \label{eq:B5}
\end{eqnarray}
where $\int_{\bd{p}} = \int \mathrm{d}^3p / (2\pi)^3$
and $B_{\rm pri}(\bd{k}_1,\bd{k}_2,\bd{k}_3) = 
M(k_1)\, M(k_2)\, M(k_3)
B_{\Phi}(\bd{k}_1,\bd{k}_2,\bd{k}_3)$ .

Previous analyses demonstrated that some terms in $B_{5}$ can dominate $B_{3}$ and $B_{4}$ at large-scale modes, while an accurate prediction of $B_{5}$ is still developing \cite{Jeong:2009vd,Sefusatti:2009qh,Karagiannis:2018jdt}.
We expect that an accurate modeling of $B_5$ needs to take into account the fact that Eq.~(\ref{eq:delta_h_exp})
does not satisfy $\langle \delta_{\rm h} \rangle = 0$.
In the case of the power spectrum calculations, 
the 1-loop corrections
require some re-normalization processes of non-linear biases 
so that the observable power spectrum can be well behaved at the limit of $\bd{k}\rightarrow\bd{0}$ \cite{McDonald:2006mx,Saito:2014aa,Desjacques:2016bnm}. 
We would need to account for similar re-normalization for $B_{5}$, but it is still uncertain.
Hence, we do not include the terms of $B_5$ when making a forecast of constraining PNG.
It would be worth noting that our analysis provides surely conservative forecasts on PNG, while it is not precise.
Nevertheless, we shall show that the post-reconstructed bispectrum analysis allows us 
to constrain PNG at 
a comparable level to the current Planck results.

\rev{On the impact of the $B_5$ terms on PNG constraints, Ref.~\cite{2020arXiv201014523M} performed likelihood analyses to infer a PNG parameter using dark matter halos in $N$-body simulations. They found that the expected best-fit PNG can be unbiased even if ignoring the $B_5$ terms in their analytic prediction of halo bispectra. Although their analysis still assumes real-space measurements, we expect that ignoring $B_5$ terms may not induce significant biased estimations of PNG when using galaxy bispetra in redshift space.}

\begin{figure}[!t]
\includegraphics[clip, width=1.0\columnwidth]{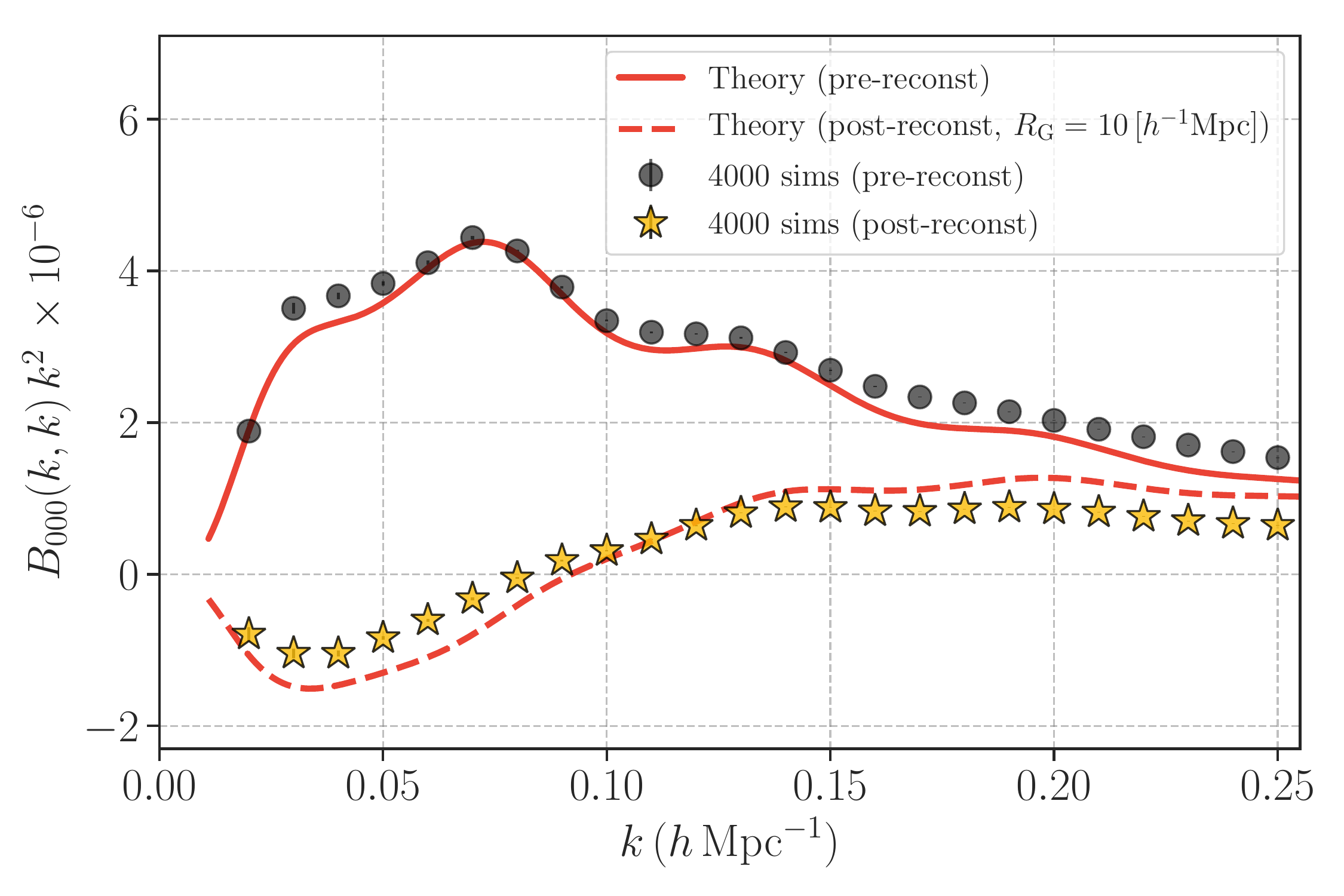}
\caption{\label{fig:fig1} Bispectrum monopole of a sample of dark matter halos with a mass range of $10^{13-14}\, h^{-1}\, M_{\odot}$ at $z=0.484$. 
The black points and star symbols represent the bispectrum monopoles at $k_{1}=k_{2}=k$
for the pre- and post-reconstructed halo density fields.
The red thick line shows the tree-level prediction by perturbation theory \cite{Scoccimarro:1999ed},
while the red dashed one stands for the tree-level prediction for the post-reconstructed density (see the details in Section~\ref{sec:model}). In this and the following figures, the bispectrum and its covariance have units of $(h^{-1} \, {\rm Mpc})^6$ and $(h^{-1} \, {\rm Mpc})^{12}$, respectively.}
\end{figure}


\begin{figure}[!t]
\includegraphics[clip, width=0.9\columnwidth]{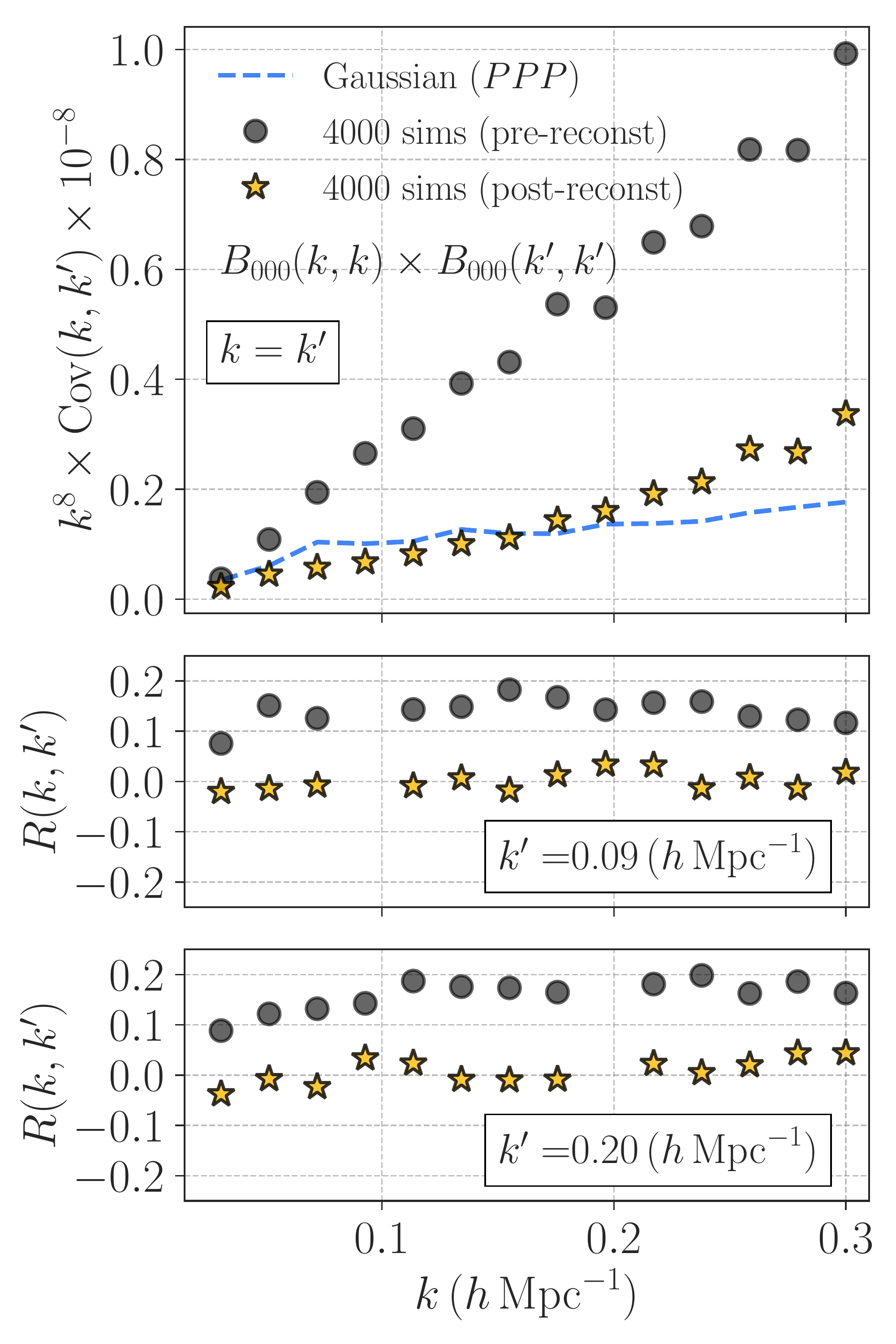}
\caption{\label{fig:fig2} The diagonal and off-diagonal elements of the bispectrum covariance for our halo sample. The top panel shows the diagonal elements in the covariance of $B_{000}(k, k)$,
while the lower two are for the off-diagonal elements. In each panel, the black points and the yellow star
symbols show the results for the pre- and post-reconstructed fields, respectively. 
For a reference, we show the Gaussian covariance by the dashed line assuming the leading-order halo power spectrum.}
\end{figure}

\begin{figure}[!t]
\includegraphics[clip, width=0.91\columnwidth]{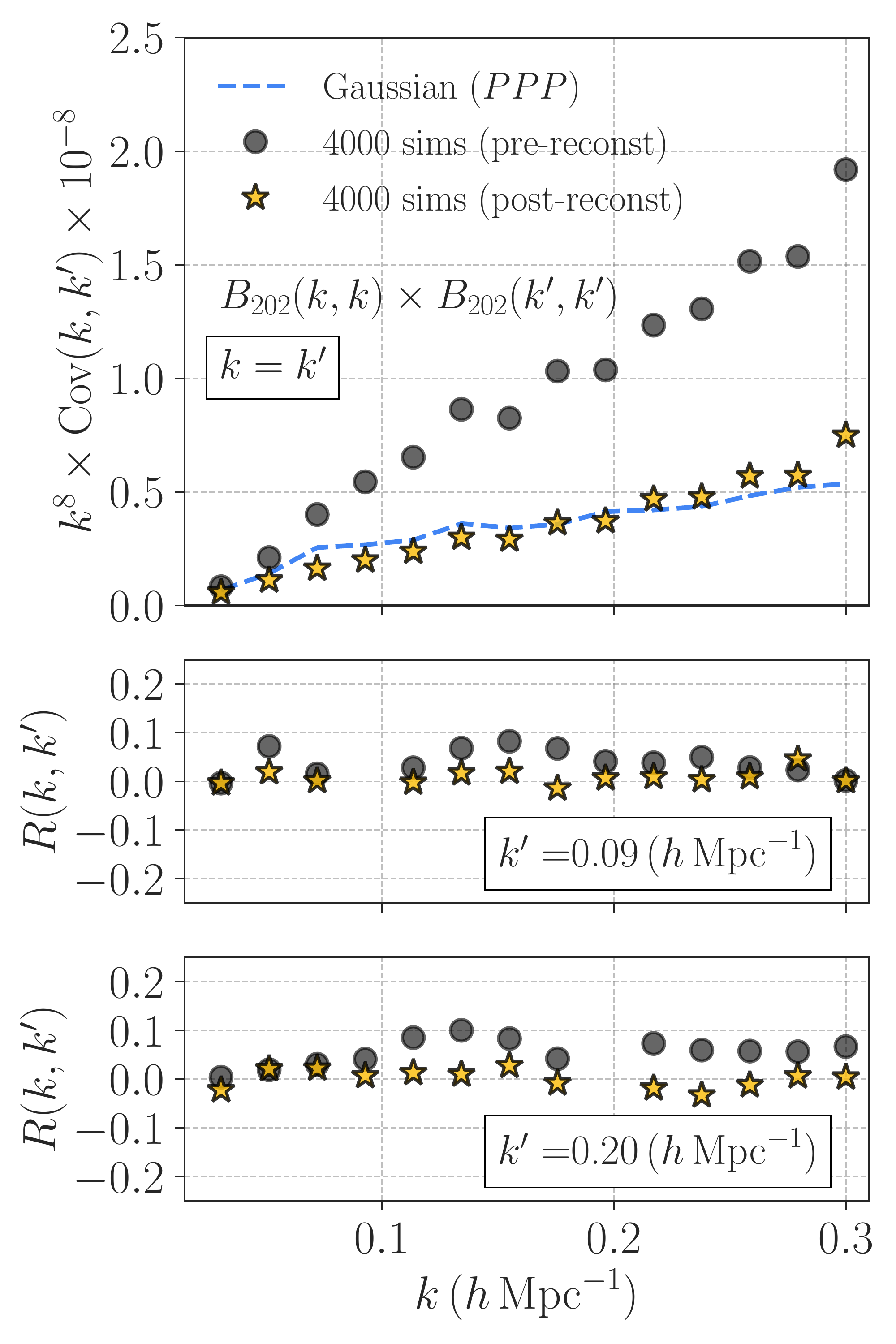}
\caption{\label{fig:fig3} Similar to Fig.~\ref{fig:fig2}, but we show the cases for $B_{202}$.}
\end{figure}

\section{Results}\label{sec:results}

\subsection{Post-reconstructed bispectrum and its covariance}

We first compare the average bispectrum monopole 
over 4000 simulations with the model prediction.
Fig.~\ref{fig:fig1} shows the comparison of $B_{000}$ at $k=k_1=k_2$ with the simulation results\footnote{For Fig.~\ref{fig:fig1}, we re-measure the bispectrum monopole of $B_{000}(k_1, k_2)$ at $k_1=k_2$ alone from 4000 simulations 
with 30 linear-spaced binning in the range of $0.01-0.3\, h\, {\rm Mpc}^{-1}$.} and 
the tree-level prediction in Eq.~(\ref{eq:B4}).
For the model prediction, 
we adopt the input value of the linear growth rate in our simulations
and set the linear bias $b_1 = 1.86$. 
We infer this linear bias by 
measuring the halo-matter cross power spectrum in real space with the simulation. 
For $b_2$, we use the fitting formula 
calibrated in Ref.~\cite{2016JCAP...02..018L}.
Assuming the tidal bias is zero at 
the initial halo density, we set $b_{K_2} = -2(b_1-1)/7$ \cite{2013PhRvD..87h3002S}.
We find that the post-reconstructed bispectrum 
can be negative at large scales and 
that our perturbative approach gives a reasonable fit to the simulation results.
\rev{We can obtain better agreements between the simulation results and our predictions at $k<0.05\, h\mathrm{Mpc}^{-1}$ in Fig.~\ref{fig:fig1} if freely varying the secondary bias parameters of $b_2$ and $b_{K_2}$.}
Ref.~\cite{Schmittfull:2015mja} has shown that the non-linear growth term in the post-reconstructed second-order matter density perturbation in real space is given by $17/21-W$. 
Hence, it is predicted to be negative ($-4/21$) on large scales in the limit of $W\to1$.
A similar argument holds even for the redshift-space halo statistics as shown in Eq.~(\ref{eq:Zrec_2}).
Note that similar negative bispectrum has been predicted for real-space matter density fields \cite{Hikage:2017tmm}.

Fig.~\ref{fig:fig2} shows the diagonal and off-diagonal elements in the covariance of $B_{000}(k,k)$
estimated by 4000 simulations. 
In the figure, we introduce the following notations of
\beqa
\mathrm{Cov}(k, k^{\prime}) &=& 
\mathrm{Cov}[B_{000}(k,k), B_{000}(k^{\prime},k^{\prime})], \\
R(k, k^{\prime}) &=& \frac{\mathrm{Cov}(k, k^{\prime})}{\left[\mathrm{Cov}(k, k)\mathrm{Cov}(k^{\prime}, k^{\prime})\right]^{1/2}},  
\eeqa
and we apply the same notation
for $B_{202}$ as well in Fig.~\ref{fig:fig3}.
We find that the covariance of the post-reconstructed bispectrum has smaller diagonal elements than the pre-reconstructed counterpart. Besides, the off-diagonal elements in the post-reconstructed bispectrum covariance becomes less prominent, allowing 
to extract nearly-independent cosmological information over different scales from the post-reconstructed bispectrum.
In Figs.~\ref{fig:fig2} and \ref{fig:fig3}, the dashed line shows a simple Gaussian covariance with the leading-order halo power spectrum based on the perturbation theory \cite{2020MNRAS.497.1684S}.
The bispectrum covariance in random density fields 
consists of four different components in general. 
One is given by the product of three power spectra, 
known as the Gaussian covariance.
\rev{We call other terms as the non-Gaussian covarinace and it consists of
\beqa
C_\mathrm{NG} \ni B^2, \, PT, \, P_6,
\eeqa
where $C_\mathrm{NG}$ is the non-Gaussian covariance of halo bispectra,
$P$, $B$, $T$, and $P_6$ are halo power spectra, bispectra, trispectra (four-point correlations in Fourier space), and six-point spectra (six-point correlations in Fourier space).
See Ref.~\cite{2020MNRAS.497.1684S} for derivations and detailed comparisons with numerical simulations.
We note that every term in the non-Gaussian covariance arises from the non-linear gravitational growth.}
By comparing the dashed line and star symbols in Figs.~\ref{fig:fig2} and \ref{fig:fig3}, 
\rev{we expect the reconstruction can suppress 
the terms of $B^2$, $T$ and $P_6$ 
in the bispectrum covariance.
At least, we confirm that the post-reconstructed bispectrum can become smaller than the pre-reconstructed counterpart in Fig~\ref{fig:fig1}.}
Although our findings in Figs.~\ref{fig:fig2} and \ref{fig:fig3} look reasonable in terms of the perturbation theory, 
more careful comparisons of the bispectrum covariance 
would be meaningful. We leave those for future studies.

\subsection{Constraining power of PNG}

\begin{figure*}[!t]
\includegraphics[clip, width=2\columnwidth]{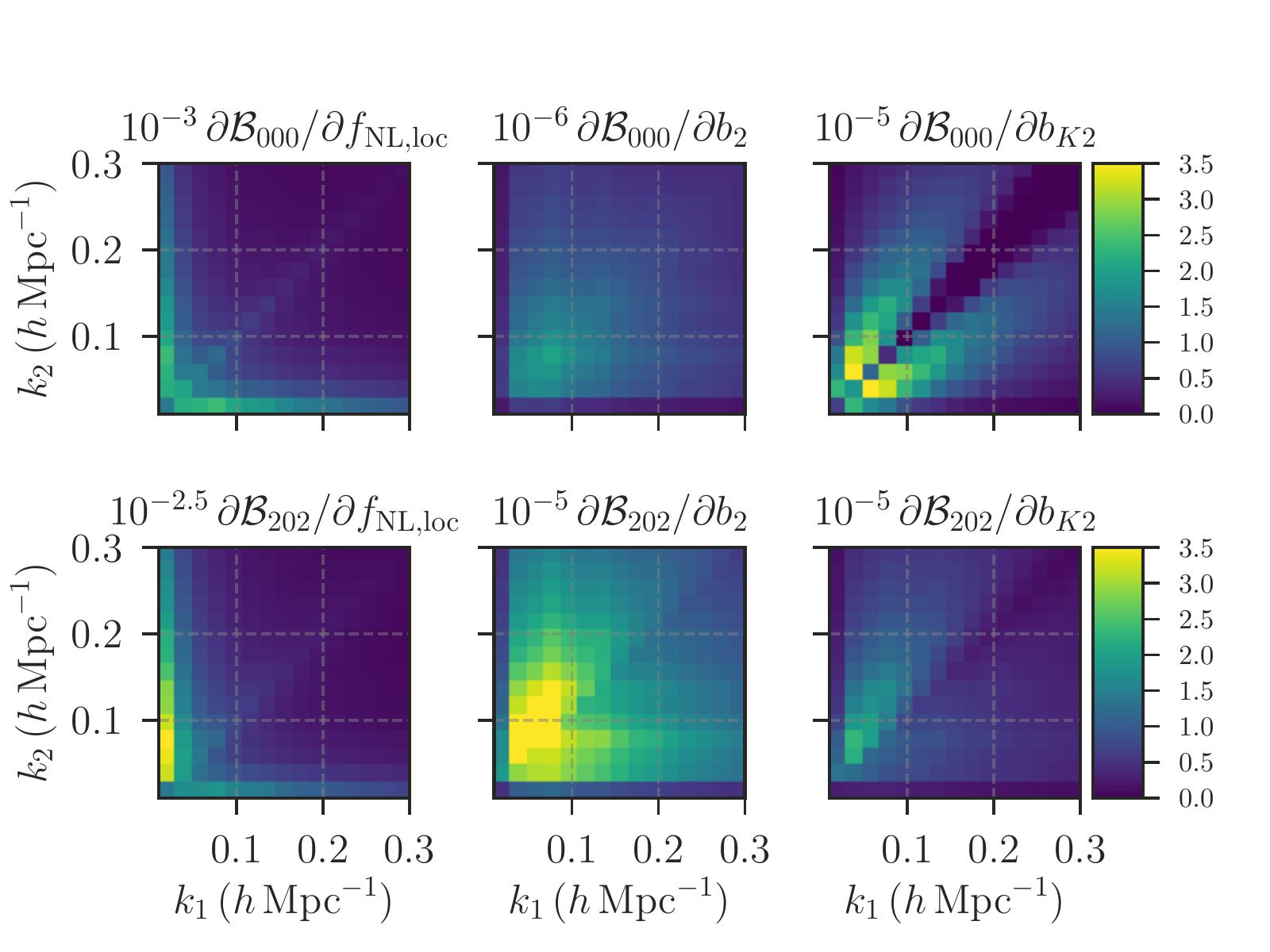}
\caption{\label{fig:fig4}
Parameter dependence of halo bispectra.
We consider a sample of dark matter halos 
with a mass range of $10^{13-14}\, h^{-1}\, M_{\odot}$ at the redshift of $0.484$.
We decompose the bispectrum 
on a basis of tri-polar spherical harmonics 
and the coefficient in the decomposition 
$B_{\ell_1 \ell_2 L}$
depends on two wave vectors $k_1$ and $k_2$ alone
(see Ref.~\cite{Sugiyama:2018yzo} for details).
We here define ${\cal B}_{\ell_1 \ell_2 L} (k_1, k_2) = k_1 k_2 B_{\ell_1 \ell_2 L}(k_1, k_2)$.
The color in each map shows the first derivative of 
${\cal B}_{000}$ or ${\cal B}_{202}$ with respect 
to parameters of $f_{\mathrm{NL,loc}}$, $b_2$ and $b_{K2}$.}
\end{figure*}

Given the result of Figs.~\ref{fig:fig2} and \ref{fig:fig3}, 
we propose to constrain PNG with the post-reconstructed galaxy bispectrum.
The reconstruction keeps the PNG-dependent galaxy bispectrum unchanged at the leading order (see Eq.~[\ref{eq:Zrec_1}]), 
while the gravity-induced bispectrum is expected to
become smaller after reconstruction as shown in Fig~\ref{fig:fig1}.
Therefore, de-correlation in galaxy-bispectrum covariance after reconstruction can provide a benefit to tightening the expected constraints of PNG for a given galaxy sample, 
compared to the case when one works with 
the pre-reconstructed density field.

To see the impact of reconstruction on constraining PNG,
we perform a Fisher analysis to study 
the expected statistical errors for 
several types of PNG. Assuming that observables follow
a multivariate Gaussian distribution, we write the Fisher matrix as
\beqa
F_{\alpha \beta} &=&
\sum_{i,j} C^{-1}_{ij} 
\frac{\partial {\cal D}_{i}}{\partial p_{\alpha}}
\frac{\partial {\cal D}_{j}}{\partial p_{\beta}}, \label{eq:Fisher_B}
\eeqa
where 
$\cal D$ represents the data vector which consists of $B_{000}$ and $B_{202}$,
$C$ is the covariance of ${\cal D}$,
$\bd{p}$ consists of the physical parameters of interest, 
and $F^{-1}$ provides the covariance matrix in parameter estimation.
In Eq.~(\ref{eq:Fisher_B}), 
the index in ${\cal D}$ is set so that two wave numbers $k_{1}$ and $k_{2}$ becomes smaller than $k_{\rm max}$.
In this paper, we adopt the following varying parameters: 
$\bd{p}= \{ f_{\rm NL}, b_2, b_{K_2}\}$
where $f_{\rm NL}$ controls the amplitude of the primordial bispectrum $B_{\rm pri}$.
We consider three different $B_{\rm pri}$, referred to as local-, equilateral-, and orthogonal-type models.
We define these three as
\beqa
B^{(\rm loc)}_{\Phi}(k_1, k_2, k_3) &=& 
2f_{\rm NL, loc} \left[ 
{\scriptstyle
P_{\Phi}(k_1) P_{\Phi}(k_2) 
+ \mbox{\small{(2 cyc.)}}
}
\right], \label{eq:f_loc}\\
B^{(\rm eq)}_{\Phi}(k_1, k_2, k_3) &=& 
-6f_{\rm NL, eq}
\Big\{
-2\left[
{\scriptstyle
P_{\Phi}(k_1) P_{\Phi}(k_2) P_{\Phi}(k_3)
}
\right]^{2/3} 
\nonumber \\
&&
+
\left[ 
{\scriptstyle 
P^{1/3}_{\Phi}(k_1) P^{2/3}_{\Phi}(k_2) P_{\Phi}(k_3)
+ \mbox{\small{(5 cyc.)}}
}
\right] 
\nonumber \\
&&
-\left[
{\scriptstyle
P_{\Phi}(k_1) P_{\Phi}(k_2) + \mbox{\small{(2 cyc.)}}
}
\right] 
\Big\}, \label{eq:f_eq} \\
B^{(\rm orth)}_{\Phi}(k_1, k_2, k_3) &=&
6f_{\rm NL, orth}\Big\{
-8\left[
{\scriptstyle
P_{\Phi}(k_1) P_{\Phi}(k_2) P_{\Phi}(k_3)
}
\right]^{2/3} 
\nonumber \\
&&
+
3\left[
{\scriptstyle
P^{1/3}_{\Phi}(k_1) P^{2/3}_{\Phi}(k_2) P_{\Phi}(k_3)
+ \mbox{\small{(5 cyc.)}}
}
\right] \nonumber \\
&&
-3\left[ 
{\scriptstyle
P_{\Phi}(k_1) P_{\Phi}(k_2) + \mbox{\small{(2 cyc.)}}
}
\right]
\Big\}, \label{eq:f_orth}
\eeqa
where 
Eqs.~(\ref{eq:f_loc})-(\ref{eq:f_orth}) represents
the local-type, equilateral-type, and orthogonal-type bispectrum, respectively.
We also study the marginalization effect of 
the non-linear biases to make the forecast of
the PNG constraints by the galaxy bispectrum.
Note that we here assume that the Kaiser factor (the kernel of $Z_1$) can be tightly constrained by power-spectrum analyses. 
When computing the Fisher matrix, we scale the covariance derived by our 4000 simulations with a survey volume of 
$4\, (h^{-1}\, {\rm Gpc})^{3}$.
This survey volume is close to the one in the BOSS.
Also, we compute the derivative terms in Eq.~(\ref{eq:Fisher_B}) by using the results of Eqs.~(\ref{eq:B3}) and (\ref{eq:B4}).
Throughout this paper, we evaluate Eq.~(\ref{eq:Fisher_B})
at $\{f_{\mathrm{NL}}, b_2, b_{K_2}\}=\{0, -0.308, -0.245\}$.

Before showing our results 
relying on
the Fisher analysis, we summarize how the bispectrum depends on the parameters of $f_{\mathrm{NL}}, b_2$ and $b_{K2}$.
Fig.~\ref{fig:fig4} shows the first derivative of $B_{000}$
or $B_{202}$ with respect to the parameters of interest.
Note that the derivatives are independent on details of reconstruction at the tree-level prediction as shown in Section~\ref{sec:model}.
Fig.~\ref{fig:fig4} highlights 
that expected parameter degeneracies
among $f_{\mathrm{NL}}$ and galaxy biases 
would be less significant for the local-type PNG.

\begin{table*}[!ht]
\begin{center}
\begin{tabular}{|c|c|c|c|c|c|}
\tableline
 & 
 $B_{000}$ (pre-reconst) &
 $B_{000}+B_{202}$ (pre-reconst)&
 $B_{000}$ (post-reconst) &
 $B_{000}+B_{202}$ (post-reconst) &
 Planck 2015
 \\ \hline
Local 
& 50.0 (45.0) & 42.4 (38.4) & 14.2 (9.65) & 13.3 (9.10) & $0.8\pm 5.0$ \\
Equilateral
& 133 (93.8) & 119 (88.3) & 97.9 (37.7) & 89.9 (35.9) & $-4 \pm 43$ \\
Orthogonal
& 79.5 (61.3) & 73.4 (57.3) & 44.8 (31.4) & 41.8 (29.8) & $-26\pm 21$
\\ \tableline
\end{tabular}
\caption{
\label{tab:fNL_const} 
Summary of the Fisher forecast of the PNG constraint by the galaxy bispectrum.
We assume the effective survey volume to be $4\,  (h^{-1}\, {\rm Gpc})^{3}$ at the redshift of $0.484$. 
We consider a sample of dark matter halos with a mass range of $10^{13-14}\, h^{-1}\, M_{\odot}$
and set the maximum wave number to be $k_{\rm max} = 0.2\, h\, {\rm Mpc}^{-1}$.
The left two columns represent the results for the pre-reconstructed field
and the right two are for the post-reconstructed field.
In each table cell, the number without brackets show the $1\sigma$ constraint of $f_{\rm NL}$
when we marginalize the second-order galaxy biases, 
while the one in brackets is 
the un-marginalized counterpart.
For comparison, we show the 68\% confidence level of $f_{\rm NL}$ provided by Planck \cite{Ade:2015ava} at the right column.
}
\end{center}
\end{table*}

The main result of this paper is shown in Table~\ref{tab:fNL_const}.
For the local-type PNG, 
we find that the post-reconstructed 
galaxy bispectrum can constrain 
$f_{\rm NL}$ with a level of 13.3
by using existing galaxy sample.
The size of error bars is still larger than the latest CMB constraint \cite{Ade:2015ava}, but it is smaller than the current best constraint by quasars \cite{2019JCAP...09..010C}.
Note that the CMB results may be subject to
biases due to secondary CMB fluctuations and cosmic infrared background \cite{2018PhRvD..98h3542H}.
In this sense, the post-reconstructed bispectrum for the BOSS galaxy sample provides a complementary probe for the PNG.
Furthermore, we demonstrate that 
the post-reconstructed bispectrum can 
improve the constraint of single PNG parameters by a factor of 1.3-3.2 compared to the original bispectrum.
For a given halo sample, 
one needs to increase the survey volume 
by a factor of 2-9
to acquire this gain without the reconstruction.

Contrary to 
what is expected from the literature,
it is worth noting 
that the gain from $B_{202}$ is not significant for the constraint of any PNG types.
In all the cases we considered, adding $B_{202}$ only improves the $f_{\mathrm{NL}}$ constraints by about $10\%$.
This indicates that one can reduce 
in practice
the degree of freedoms by using $B_{000}$ alone 
for constraining PNG.

\begin{figure}[!t]
\includegraphics[clip, width=0.95\columnwidth]{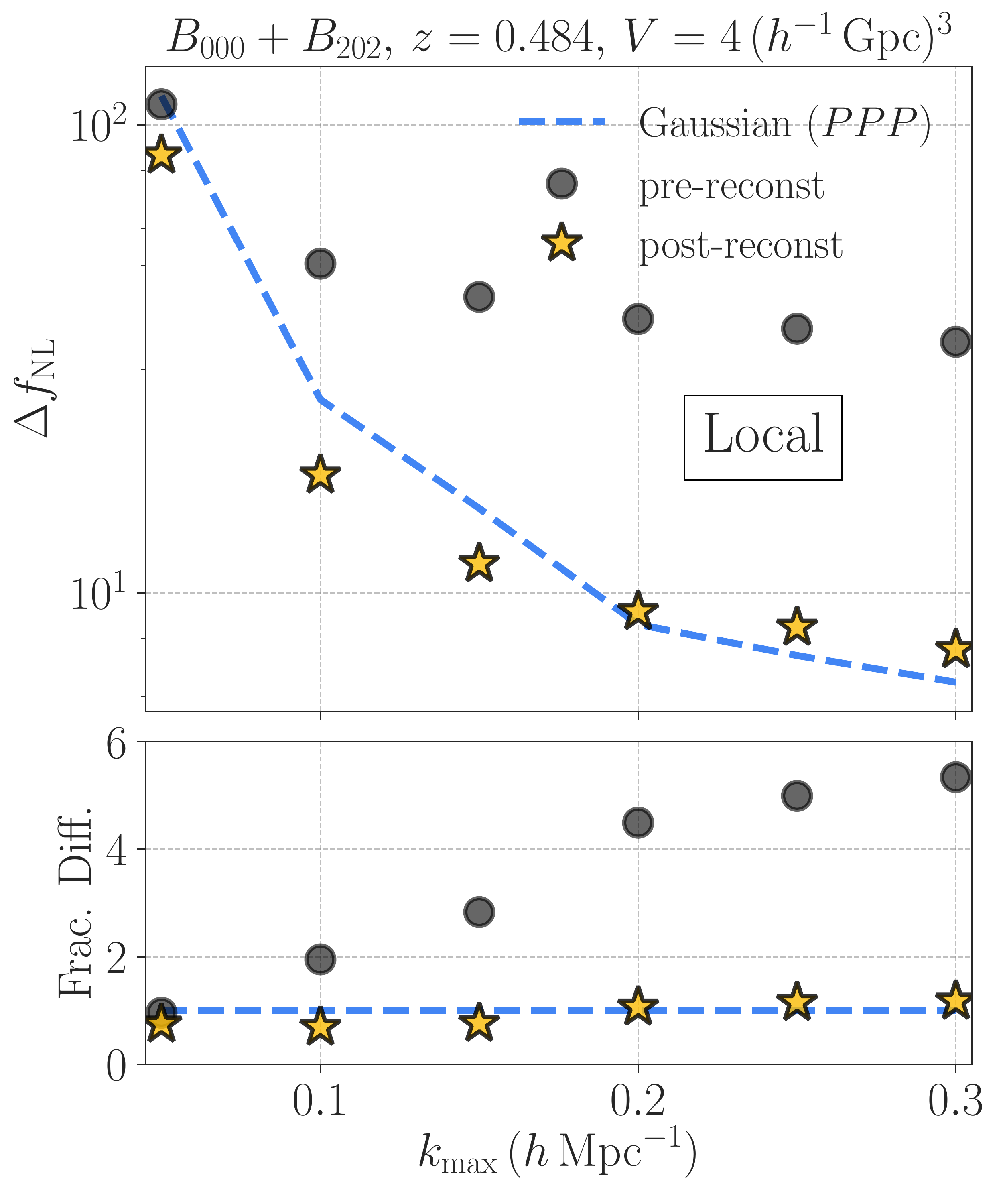}
\caption{\label{fig:fig5} 
Un-marginalized one-sigma confidence level of 
the local-type PNG $f_{\rm NL, loc}$ for different maximum wave numbers $k_{\mathrm{max}}$ in bispectrum analyses.
In this figure, we assume the effective survey volume to be $4\,  (h^{-1}\, {\rm Gpc})^{3}$ at the redshift of $0.484$. We consider a sample of dark matter halos with a mass range of $10^{13-14}\, h^{-1}\, M_{\odot}$.
The black points show the results for the pre-reconstructed bispectrum, while the yellow star symbols represent the post-reconstructed cases.
For a comparison, we show the results based on the Gaussian covariance by the cyan dashed line.
}
\end{figure}

Finally, we study 
the un-marginalized error of $f_{\mathrm{NL}}$ for different maximum wave vectors $k_{\mathrm{max}}$
to clarify the effect of the reconstruction on 
the bispectrum covariance.
Fig.~\ref{fig:fig5} shows the un-marginalized errors for the local-type PNG as a function of $k_{\mathrm{max}}$.
The comparison between 
black points and the yellow star symbols in the figure shows
that the reconstruction becomes efficient to reduce the off-diagonal bispectrum covariances at $k\simgt0.1\, h\, \mathrm{Mpc}^{-1}$.
The post-reconstructed results closely follow the Gaussian-covariance expectations, but there exist substantial differences at weakly non-linear scales of 
$k=0.1-0.15\, h\, \mathrm{Mpc}^{-1}$.
After some trials, we found that 
these differences can be caused by 
negative off-diagonal 
covariances of the post-reconstructed bispectrum.
Similar trends have been found in Ref.~\cite{2020arXiv200713998H} 
for the post-reconstructed matter power spectrum.
The results in Fig.~\ref{fig:fig5} indicate that one will be able to design an optimal reconstruction so that the error of $f_{\mathrm{NL}}$ can be minimized for a given $k_{\mathrm{max}}$.
Such optimizations are of great interest but beyond the scope of this paper.

\section{Conclusion and Discussions}\label{sec:con}

The galaxy bispectrum represents an interesting probe of inflationary physics in the early Universe, allowing to measure various types of PNG. 
The numerical calculations presented in this work show that 
the same algorithm used for BAO reconstruction as introduced 
in Ref.~\cite{Eisenstein:2006nk} can become an essential tool to 
achieve the expected accuracy in PNG constraints 
from future galaxy surveys aimed by the science community.
As a representative example, 
assuming the SDSS-III BOSS galaxy sample, 
we found that the galaxy bispectrum under the realistic non-Gaussian covariance can constrain the PNG with a level of $\Delta f_{\rm NL}=42.4$, $119$ and $73.4$ for 
the local-, equilateral- and orthogonal-type models, respectively.
Nevertheless, the post-reconstructed bispectrum 
can improve this constraint by a factor of 1.3-3.2
when one restricts the measurements to be in quasi-linear scales ($k\le 0.2\, h\, {\rm Mpc}^{-1}$).
We here emphasize that our Fisher analyses 
do not include the 1-loop corrections of the bispectrum (Eq.~[\ref{eq:B5}]), providing a surely conservative forecast of the PNG constraints.

So far we have considered the standard BAO reconstruction applied to 
massive halos corresponding 
to luminous red galaxy like objects.
We expect stronger PNG constraints from higher number densities going to lower mass halos, 
corresponding to emission-line galaxies, 
as will be detected by future galaxy surveys, 
e.g. EUCLID\footnote{\url{https://sci.esa.int/web/euclid}} 
or DESI\footnote{\url{https://www.desi.lbl.gov/}}. 
The approach proposed in this study will 
be even more crucial to extract PNG signatures, 
as such surveys provide data tracing further the non-linear regime of the cosmic density field.
However, there still remain important issues to be resolved before we apply our proposal to real data sets.
For instance, we need an accurate modeling for 
the 1-loop-correction terms $B_5$ as well as some corrections for mode-mixing effects by a complex survey window.
We also found that the anisotropic bispectrum signal $B_{202}$ would not be relevant to improve the PNG constraints.
Nevertheless, it is still beneficial to study higher-order terms in the monopole bispectrum such as $B_{110}$ and $B_{220}$ for further improvements in the PNG constraints.

In summary, this work represents a new approach to investigate PNG from the large scale structure. 
A lot of work still needs to be done following this path.
In a forthcoming paper, we will study the anisotropic signals of the post-reconstructed bispectrum and present the importance of reconstruction to optimize the redshift-space analysis 
of the galaxy bispectrum.

\begin{acknowledgments}
We thank Shun Saito, Florian Beutler and Hee-Jong Seo for useful comments.
This work is supported by MEXT KAKENHI
Grant Number (15H05893, 17H01131, 18H04358, 19K14767, 20H04723).
NSS acknowledges financial support from JSPS KAKENHI Grant Number 19K14703.
FSK thanks support from grants SEV-2015-0548, RYC2015-18693
and AYA2017-89891-P.
Numerical computations were carried out on Cray XC50 
at the Center for Computational Astrophysics in NAOJ.
\end{acknowledgments}



\bibliography{apssamp}

\begin{thebibliography}{42}%
\makeatletter
\providecommand \@ifxundefined [1]{%
 \@ifx{#1\undefined}
}%
\providecommand \@ifnum [1]{%
 \ifnum #1\expandafter \@firstoftwo
 \else \expandafter \@secondoftwo
 \fi
}%
\providecommand \@ifx [1]{%
 \ifx #1\expandafter \@firstoftwo
 \else \expandafter \@secondoftwo
 \fi
}%
\providecommand \natexlab [1]{#1}%
\providecommand \enquote  [1]{``#1''}%
\providecommand \bibnamefont  [1]{#1}%
\providecommand \bibfnamefont [1]{#1}%
\providecommand \citenamefont [1]{#1}%
\providecommand \href@noop [0]{\@secondoftwo}%
\providecommand \href [0]{\begingroup \@sanitize@url \@href}%
\providecommand \@href[1]{\@@startlink{#1}\@@href}%
\providecommand \@@href[1]{\endgroup#1\@@endlink}%
\providecommand \@sanitize@url [0]{\catcode `\\12\catcode `\$12\catcode
  `\&12\catcode `\#12\catcode `\^12\catcode `\_12\catcode `\%12\relax}%
\providecommand \@@startlink[1]{}%
\providecommand \@@endlink[0]{}%
\providecommand \url  [0]{\begingroup\@sanitize@url \@url }%
\providecommand \@url [1]{\endgroup\@href {#1}{\urlprefix }}%
\providecommand \urlprefix  [0]{URL }%
\providecommand \Eprint [0]{\href }%
\providecommand \doibase [0]{https://doi.org/}%
\providecommand \selectlanguage [0]{\@gobble}%
\providecommand \bibinfo  [0]{\@secondoftwo}%
\providecommand \bibfield  [0]{\@secondoftwo}%
\providecommand \translation [1]{[#1]}%
\providecommand \BibitemOpen [0]{}%
\providecommand \bibitemStop [0]{}%
\providecommand \bibitemNoStop [0]{.\EOS\space}%
\providecommand \EOS [0]{\spacefactor3000\relax}%
\providecommand \BibitemShut  [1]{\csname bibitem#1\endcsname}%
\let\auto@bib@innerbib\@empty
\bibitem [{\citenamefont {Bartolo}\ \emph {et~al.}(2004)\citenamefont
  {Bartolo}, \citenamefont {Komatsu}, \citenamefont {Matarrese},\ and\
  \citenamefont {Riotto}}]{Bartolo:2004if}%
  \BibitemOpen
  \bibfield  {author} {\bibinfo {author} {\bibfnamefont {N.}~\bibnamefont
  {Bartolo}}, \bibinfo {author} {\bibfnamefont {E.}~\bibnamefont {Komatsu}},
  \bibinfo {author} {\bibfnamefont {S.}~\bibnamefont {Matarrese}},\ and\
  \bibinfo {author} {\bibfnamefont {A.}~\bibnamefont {Riotto}},\ }\href
  {https://doi.org/10.1016/j.physrep.2004.08.022} {\bibfield  {journal}
  {\bibinfo  {journal} {Phys. Rept.}\ }\textbf {\bibinfo {volume} {402}},\
  \bibinfo {pages} {103} (\bibinfo {year} {2004})},\ \Eprint
  {https://arxiv.org/abs/astro-ph/0406398} {arXiv:astro-ph/0406398 [astro-ph]}
  \BibitemShut {NoStop}%
\bibitem [{\citenamefont {Komatsu}\ \emph {et~al.}(2005)\citenamefont
  {Komatsu}, \citenamefont {Spergel},\ and\ \citenamefont
  {Wandelt}}]{Komatsu:2003iq}%
  \BibitemOpen
  \bibfield  {author} {\bibinfo {author} {\bibfnamefont {E.}~\bibnamefont
  {Komatsu}}, \bibinfo {author} {\bibfnamefont {D.~N.}\ \bibnamefont
  {Spergel}},\ and\ \bibinfo {author} {\bibfnamefont {B.~D.}\ \bibnamefont
  {Wandelt}},\ }\href {https://doi.org/10.1086/491724} {\bibfield  {journal}
  {\bibinfo  {journal} {Astrophys. J.}\ }\textbf {\bibinfo {volume} {634}},\
  \bibinfo {pages} {14} (\bibinfo {year} {2005})},\ \Eprint
  {https://arxiv.org/abs/astro-ph/0305189} {arXiv:astro-ph/0305189 [astro-ph]}
  \BibitemShut {NoStop}%
\bibitem [{\citenamefont {Sefusatti}\ and\ \citenamefont
  {Komatsu}(2007)}]{Sefusatti:2007ih}%
  \BibitemOpen
  \bibfield  {author} {\bibinfo {author} {\bibfnamefont {E.}~\bibnamefont
  {Sefusatti}}\ and\ \bibinfo {author} {\bibfnamefont {E.}~\bibnamefont
  {Komatsu}},\ }\href {https://doi.org/10.1103/PhysRevD.76.083004} {\bibfield
  {journal} {\bibinfo  {journal} {Phys. Rev.}\ }\textbf {\bibinfo {volume}
  {D76}},\ \bibinfo {pages} {083004} (\bibinfo {year} {2007})},\ \Eprint
  {https://arxiv.org/abs/0705.0343} {arXiv:0705.0343 [astro-ph]} \BibitemShut
  {NoStop}%
\bibitem [{\citenamefont {Ade}\ \emph {et~al.}(2016{\natexlab{a}})\citenamefont
  {Ade} \emph {et~al.}}]{Ade:2015ava}%
  \BibitemOpen
  \bibfield  {author} {\bibinfo {author} {\bibfnamefont {P.~A.~R.}\
  \bibnamefont {Ade}} \emph {et~al.} (\bibinfo {collaboration} {Planck}),\
  }\href {https://doi.org/10.1051/0004-6361/201525836} {\bibfield  {journal}
  {\bibinfo  {journal} {Astron. Astrophys.}\ }\textbf {\bibinfo {volume}
  {594}},\ \bibinfo {pages} {A17} (\bibinfo {year} {2016}{\natexlab{a}})},\
  \Eprint {https://arxiv.org/abs/1502.01592} {arXiv:1502.01592 [astro-ph.CO]}
  \BibitemShut {NoStop}%
\bibitem [{\citenamefont {Karagiannis}\ \emph {et~al.}(2018)\citenamefont
  {Karagiannis}, \citenamefont {Lazanu}, \citenamefont {Liguori}, \citenamefont
  {Raccanelli}, \citenamefont {Bartolo},\ and\ \citenamefont
  {Verde}}]{Karagiannis:2018jdt}%
  \BibitemOpen
  \bibfield  {author} {\bibinfo {author} {\bibfnamefont {D.}~\bibnamefont
  {Karagiannis}}, \bibinfo {author} {\bibfnamefont {A.}~\bibnamefont {Lazanu}},
  \bibinfo {author} {\bibfnamefont {M.}~\bibnamefont {Liguori}}, \bibinfo
  {author} {\bibfnamefont {A.}~\bibnamefont {Raccanelli}}, \bibinfo {author}
  {\bibfnamefont {N.}~\bibnamefont {Bartolo}},\ and\ \bibinfo {author}
  {\bibfnamefont {L.}~\bibnamefont {Verde}},\ }\href
  {https://doi.org/10.1093/mnras/sty1029} {\bibfield  {journal} {\bibinfo
  {journal} {Mon. Not. Roy. Astron. Soc.}\ }\textbf {\bibinfo {volume} {478}},\
  \bibinfo {pages} {1341} (\bibinfo {year} {2018})},\ \Eprint
  {https://arxiv.org/abs/1801.09280} {arXiv:1801.09280 [astro-ph.CO]}
  \BibitemShut {NoStop}%
\bibitem [{\citenamefont {{Ferraro}}\ and\ \citenamefont
  {{Wilson}}(2019)}]{2019BAAS...51c..72F}%
  \BibitemOpen
  \bibfield  {author} {\bibinfo {author} {\bibfnamefont {S.}~\bibnamefont
  {{Ferraro}}}\ and\ \bibinfo {author} {\bibfnamefont {M.~J.}\ \bibnamefont
  {{Wilson}}},\ }\href@noop {} {\bibfield  {journal} {\bibinfo  {journal}
  {\baas}\ }\textbf {\bibinfo {volume} {51}},\ \bibinfo {eid} {72} (\bibinfo
  {year} {2019})},\ \Eprint {https://arxiv.org/abs/1903.09208}
  {arXiv:1903.09208 [astro-ph.CO]} \BibitemShut {NoStop}%
\bibitem [{\citenamefont {{Gualdi}}\ and\ \citenamefont
  {{Verde}}(2020)}]{2020JCAP...06..041G}%
  \BibitemOpen
  \bibfield  {author} {\bibinfo {author} {\bibfnamefont {D.}~\bibnamefont
  {{Gualdi}}}\ and\ \bibinfo {author} {\bibfnamefont {L.}~\bibnamefont
  {{Verde}}},\ }\href {https://doi.org/10.1088/1475-7516/2020/06/041}
  {\bibfield  {journal} {\bibinfo  {journal} {\jcap}\ }\textbf {\bibinfo
  {volume} {2020}},\ \bibinfo {eid} {041} (\bibinfo {year} {2020})},\ \Eprint
  {https://arxiv.org/abs/2003.12075} {arXiv:2003.12075 [astro-ph.CO]}
  \BibitemShut {NoStop}%
\bibitem [{\citenamefont {Chan}\ and\ \citenamefont
  {Blot}(2017)}]{Chan:2016ehg}%
  \BibitemOpen
  \bibfield  {author} {\bibinfo {author} {\bibfnamefont {K.~C.}\ \bibnamefont
  {Chan}}\ and\ \bibinfo {author} {\bibfnamefont {L.}~\bibnamefont {Blot}},\
  }\href {https://doi.org/10.1103/PhysRevD.96.023528} {\bibfield  {journal}
  {\bibinfo  {journal} {Phys. Rev.}\ }\textbf {\bibinfo {volume} {D96}},\
  \bibinfo {pages} {023528} (\bibinfo {year} {2017})},\ \Eprint
  {https://arxiv.org/abs/1610.06585} {arXiv:1610.06585 [astro-ph.CO]}
  \BibitemShut {NoStop}%
\bibitem [{\citenamefont {{Sugiyama}}\ \emph {et~al.}(2020)\citenamefont
  {{Sugiyama}}, \citenamefont {{Saito}}, \citenamefont {{Beutler}},\ and\
  \citenamefont {{Seo}}}]{2020MNRAS.497.1684S}%
  \BibitemOpen
  \bibfield  {author} {\bibinfo {author} {\bibfnamefont {N.~S.}\ \bibnamefont
  {{Sugiyama}}}, \bibinfo {author} {\bibfnamefont {S.}~\bibnamefont {{Saito}}},
  \bibinfo {author} {\bibfnamefont {F.}~\bibnamefont {{Beutler}}},\ and\
  \bibinfo {author} {\bibfnamefont {H.-J.}\ \bibnamefont {{Seo}}},\ }\href
  {https://doi.org/10.1093/mnras/staa1940} {\bibfield  {journal} {\bibinfo
  {journal} {\mnras}\ }\textbf {\bibinfo {volume} {497}},\ \bibinfo {pages}
  {1684} (\bibinfo {year} {2020})},\ \Eprint {https://arxiv.org/abs/1908.06234}
  {arXiv:1908.06234 [astro-ph.CO]} \BibitemShut {NoStop}%
\bibitem [{\citenamefont {Eisenstein}\ \emph {et~al.}(2007)\citenamefont
  {Eisenstein}, \citenamefont {Seo}, \citenamefont {Sirko},\ and\ \citenamefont
  {Spergel}}]{Eisenstein:2006nk}%
  \BibitemOpen
  \bibfield  {author} {\bibinfo {author} {\bibfnamefont {D.~J.}\ \bibnamefont
  {Eisenstein}}, \bibinfo {author} {\bibfnamefont {H.-j.}\ \bibnamefont {Seo}},
  \bibinfo {author} {\bibfnamefont {E.}~\bibnamefont {Sirko}},\ and\ \bibinfo
  {author} {\bibfnamefont {D.}~\bibnamefont {Spergel}},\ }\href
  {https://doi.org/10.1086/518712} {\bibfield  {journal} {\bibinfo  {journal}
  {Astrophys. J.}\ }\textbf {\bibinfo {volume} {664}},\ \bibinfo {pages} {675}
  (\bibinfo {year} {2007})},\ \Eprint {https://arxiv.org/abs/astro-ph/0604362}
  {arXiv:astro-ph/0604362 [astro-ph]} \BibitemShut {NoStop}%
\bibitem [{\citenamefont {{Slepian}}\ \emph {et~al.}(2017)\citenamefont
  {{Slepian}}, \citenamefont {{Eisenstein}}, \citenamefont {{Brownstein}},
  \citenamefont {{Chuang}}, \citenamefont {{Gil-Mar{\'\i}n}}, \citenamefont
  {{Ho}}, \citenamefont {{Kitaura}}, \citenamefont {{Percival}}, \citenamefont
  {{Ross}}, \citenamefont {{Rossi}}, \citenamefont {{Seo}}, \citenamefont
  {{Slosar}},\ and\ \citenamefont {{Vargas-Maga{\~n}a}}}]{2017MNRAS.469.1738S}%
  \BibitemOpen
  \bibfield  {author} {\bibinfo {author} {\bibfnamefont {Z.}~\bibnamefont
  {{Slepian}}}, \bibinfo {author} {\bibfnamefont {D.~J.}\ \bibnamefont
  {{Eisenstein}}}, \bibinfo {author} {\bibfnamefont {J.~R.}\ \bibnamefont
  {{Brownstein}}}, \bibinfo {author} {\bibfnamefont {C.-H.}\ \bibnamefont
  {{Chuang}}}, \bibinfo {author} {\bibfnamefont {H.}~\bibnamefont
  {{Gil-Mar{\'\i}n}}}, \bibinfo {author} {\bibfnamefont {S.}~\bibnamefont
  {{Ho}}}, \bibinfo {author} {\bibfnamefont {F.-S.}\ \bibnamefont {{Kitaura}}},
  \bibinfo {author} {\bibfnamefont {W.~J.}\ \bibnamefont {{Percival}}},
  \bibinfo {author} {\bibfnamefont {A.~J.}\ \bibnamefont {{Ross}}}, \bibinfo
  {author} {\bibfnamefont {G.}~\bibnamefont {{Rossi}}}, \bibinfo {author}
  {\bibfnamefont {H.-J.}\ \bibnamefont {{Seo}}}, \bibinfo {author}
  {\bibfnamefont {A.}~\bibnamefont {{Slosar}}},\ and\ \bibinfo {author}
  {\bibfnamefont {M.}~\bibnamefont {{Vargas-Maga{\~n}a}}},\ }\href
  {https://doi.org/10.1093/mnras/stx488} {\bibfield  {journal} {\bibinfo
  {journal} {\mnras}\ }\textbf {\bibinfo {volume} {469}},\ \bibinfo {pages}
  {1738} (\bibinfo {year} {2017})},\ \Eprint {https://arxiv.org/abs/1607.06097}
  {arXiv:1607.06097 [astro-ph.CO]} \BibitemShut {NoStop}%
\bibitem [{\citenamefont {Padmanabhan}\ \emph {et~al.}(2009)\citenamefont
  {Padmanabhan}, \citenamefont {White},\ and\ \citenamefont
  {Cohn}}]{Padmanabhan:2008dd}%
  \BibitemOpen
  \bibfield  {author} {\bibinfo {author} {\bibfnamefont {N.}~\bibnamefont
  {Padmanabhan}}, \bibinfo {author} {\bibfnamefont {M.}~\bibnamefont {White}},\
  and\ \bibinfo {author} {\bibfnamefont {J.~D.}\ \bibnamefont {Cohn}},\ }\href
  {https://doi.org/10.1103/PhysRevD.79.063523} {\bibfield  {journal} {\bibinfo
  {journal} {Phys. Rev.}\ }\textbf {\bibinfo {volume} {D79}},\ \bibinfo {pages}
  {063523} (\bibinfo {year} {2009})},\ \Eprint
  {https://arxiv.org/abs/0812.2905} {arXiv:0812.2905 [astro-ph]} \BibitemShut
  {NoStop}%
\bibitem [{\citenamefont {{Hikage}}\ \emph {et~al.}(2020)\citenamefont
  {{Hikage}}, \citenamefont {{Takahashi}},\ and\ \citenamefont
  {{Koyama}}}]{2020arXiv200713998H}%
  \BibitemOpen
  \bibfield  {author} {\bibinfo {author} {\bibfnamefont {C.}~\bibnamefont
  {{Hikage}}}, \bibinfo {author} {\bibfnamefont {R.}~\bibnamefont
  {{Takahashi}}},\ and\ \bibinfo {author} {\bibfnamefont {K.}~\bibnamefont
  {{Koyama}}},\ }\href@noop {} {\bibfield  {journal} {\bibinfo  {journal}
  {arXiv e-prints}\ ,\ \bibinfo {eid} {arXiv:2007.13998}} (\bibinfo {year}
  {2020})},\ \Eprint {https://arxiv.org/abs/2007.13998} {arXiv:2007.13998
  [astro-ph.CO]} \BibitemShut {NoStop}%
\bibitem [{\citenamefont {Sugiyama}\ \emph {et~al.}(2019)\citenamefont
  {Sugiyama}, \citenamefont {Saito}, \citenamefont {Beutler},\ and\
  \citenamefont {Seo}}]{Sugiyama:2018yzo}%
  \BibitemOpen
  \bibfield  {author} {\bibinfo {author} {\bibfnamefont {N.~S.}\ \bibnamefont
  {Sugiyama}}, \bibinfo {author} {\bibfnamefont {S.}~\bibnamefont {Saito}},
  \bibinfo {author} {\bibfnamefont {F.}~\bibnamefont {Beutler}},\ and\ \bibinfo
  {author} {\bibfnamefont {H.-J.}\ \bibnamefont {Seo}},\ }\href
  {https://doi.org/10.1093/mnras/sty3249} {\bibfield  {journal} {\bibinfo
  {journal} {Mon. Not. Roy. Astron. Soc.}\ }\textbf {\bibinfo {volume} {484}},\
  \bibinfo {pages} {364} (\bibinfo {year} {2019})},\ \Eprint
  {https://arxiv.org/abs/1803.02132} {arXiv:1803.02132 [astro-ph.CO]}
  \BibitemShut {NoStop}%
\bibitem [{\citenamefont {Springel}(2005)}]{Springel:2005mi}%
  \BibitemOpen
  \bibfield  {author} {\bibinfo {author} {\bibfnamefont {V.}~\bibnamefont
  {Springel}},\ }\href {https://doi.org/10.1111/j.1365-2966.2005.09655.x}
  {\bibfield  {journal} {\bibinfo  {journal} {Mon. Not. Roy. Astron. Soc.}\
  }\textbf {\bibinfo {volume} {364}},\ \bibinfo {pages} {1105} (\bibinfo {year}
  {2005})},\ \Eprint {https://arxiv.org/abs/astro-ph/0505010}
  {arXiv:astro-ph/0505010 [astro-ph]} \BibitemShut {NoStop}%
\bibitem [{\citenamefont {{Nishimichi}}\ \emph {et~al.}(2009)\citenamefont
  {{Nishimichi}}, \citenamefont {{Shirata}}, \citenamefont {{Taruya}},
  \citenamefont {{Yahata}}, \citenamefont {{Saito}}, \citenamefont {{Suto}},
  \citenamefont {{Takahashi}}, \citenamefont {{Yoshida}}, \citenamefont
  {{Matsubara}}, \citenamefont {{Sugiyama}}, \citenamefont {{Kayo}},
  \citenamefont {{Jing}},\ and\ \citenamefont
  {{Yoshikawa}}}]{2009PASJ...61..321N}%
  \BibitemOpen
  \bibfield  {author} {\bibinfo {author} {\bibfnamefont {T.}~\bibnamefont
  {{Nishimichi}}}, \bibinfo {author} {\bibfnamefont {A.}~\bibnamefont
  {{Shirata}}}, \bibinfo {author} {\bibfnamefont {A.}~\bibnamefont {{Taruya}}},
  \bibinfo {author} {\bibfnamefont {K.}~\bibnamefont {{Yahata}}}, \bibinfo
  {author} {\bibfnamefont {S.}~\bibnamefont {{Saito}}}, \bibinfo {author}
  {\bibfnamefont {Y.}~\bibnamefont {{Suto}}}, \bibinfo {author} {\bibfnamefont
  {R.}~\bibnamefont {{Takahashi}}}, \bibinfo {author} {\bibfnamefont
  {N.}~\bibnamefont {{Yoshida}}}, \bibinfo {author} {\bibfnamefont
  {T.}~\bibnamefont {{Matsubara}}}, \bibinfo {author} {\bibfnamefont
  {N.}~\bibnamefont {{Sugiyama}}}, \bibinfo {author} {\bibfnamefont
  {I.}~\bibnamefont {{Kayo}}}, \bibinfo {author} {\bibfnamefont
  {Y.}~\bibnamefont {{Jing}}},\ and\ \bibinfo {author} {\bibfnamefont
  {K.}~\bibnamefont {{Yoshikawa}}},\ }\href
  {https://doi.org/10.1093/pasj/61.2.321} {\bibfield  {journal} {\bibinfo
  {journal} {Publications of the Astronomical Society of Japan}\ }\textbf
  {\bibinfo {volume} {61}},\ \bibinfo {pages} {321} (\bibinfo {year} {2009})},\
  \Eprint {https://arxiv.org/abs/0810.0813} {arXiv:0810.0813 [astro-ph]}
  \BibitemShut {NoStop}%
\bibitem [{\citenamefont {{Valageas}}\ and\ \citenamefont
  {{Nishimichi}}(2011)}]{2011A&A...527A..87V}%
  \BibitemOpen
  \bibfield  {author} {\bibinfo {author} {\bibfnamefont {P.}~\bibnamefont
  {{Valageas}}}\ and\ \bibinfo {author} {\bibfnamefont {T.}~\bibnamefont
  {{Nishimichi}}},\ }\href {https://doi.org/10.1051/0004-6361/201015685}
  {\bibfield  {journal} {\bibinfo  {journal} {\aap}\ }\textbf {\bibinfo
  {volume} {527}},\ \bibinfo {eid} {A87} (\bibinfo {year} {2011})},\ \Eprint
  {https://arxiv.org/abs/1009.0597} {arXiv:1009.0597 [astro-ph.CO]}
  \BibitemShut {NoStop}%
\bibitem [{\citenamefont {Crocce}\ \emph {et~al.}(2006)\citenamefont {Crocce},
  \citenamefont {Pueblas},\ and\ \citenamefont {Scoccimarro}}]{Crocce:2006ve}%
  \BibitemOpen
  \bibfield  {author} {\bibinfo {author} {\bibfnamefont {M.}~\bibnamefont
  {Crocce}}, \bibinfo {author} {\bibfnamefont {S.}~\bibnamefont {Pueblas}},\
  and\ \bibinfo {author} {\bibfnamefont {R.}~\bibnamefont {Scoccimarro}},\
  }\href {https://doi.org/10.1111/j.1365-2966.2006.11040.x} {\bibfield
  {journal} {\bibinfo  {journal} {Mon. Not. Roy. Astron. Soc.}\ }\textbf
  {\bibinfo {volume} {373}},\ \bibinfo {pages} {369} (\bibinfo {year}
  {2006})},\ \Eprint {https://arxiv.org/abs/astro-ph/0606505}
  {arXiv:astro-ph/0606505 [astro-ph]} \BibitemShut {NoStop}%
\bibitem [{\citenamefont {Lewis}\ \emph {et~al.}(2000)\citenamefont {Lewis},
  \citenamefont {Challinor},\ and\ \citenamefont {Lasenby}}]{Lewis:1999bs}%
  \BibitemOpen
  \bibfield  {author} {\bibinfo {author} {\bibfnamefont {A.}~\bibnamefont
  {Lewis}}, \bibinfo {author} {\bibfnamefont {A.}~\bibnamefont {Challinor}},\
  and\ \bibinfo {author} {\bibfnamefont {A.}~\bibnamefont {Lasenby}},\ }\href
  {https://doi.org/10.1086/309179} {\bibfield  {journal} {\bibinfo  {journal}
  {\apj}\ }\textbf {\bibinfo {volume} {538}},\ \bibinfo {pages} {473} (\bibinfo
  {year} {2000})},\ \Eprint {https://arxiv.org/abs/astro-ph/9911177}
  {arXiv:astro-ph/9911177 [astro-ph]} \BibitemShut {NoStop}%
\bibitem [{\citenamefont {Ade}\ \emph {et~al.}(2016{\natexlab{b}})\citenamefont
  {Ade} \emph {et~al.}}]{Ade:2015xua}%
  \BibitemOpen
  \bibfield  {author} {\bibinfo {author} {\bibfnamefont {P.~A.~R.}\
  \bibnamefont {Ade}} \emph {et~al.} (\bibinfo {collaboration} {Planck}),\
  }\href {https://doi.org/10.1051/0004-6361/201525830} {\bibfield  {journal}
  {\bibinfo  {journal} {Astron. Astrophys.}\ }\textbf {\bibinfo {volume}
  {594}},\ \bibinfo {pages} {A13} (\bibinfo {year} {2016}{\natexlab{b}})},\
  \Eprint {https://arxiv.org/abs/1502.01589} {arXiv:1502.01589 [astro-ph.CO]}
  \BibitemShut {NoStop}%
\bibitem [{\citenamefont {Dawson}\ \emph {et~al.}(2013)\citenamefont {Dawson}
  \emph {et~al.}}]{Dawson:2012va}%
  \BibitemOpen
  \bibfield  {author} {\bibinfo {author} {\bibfnamefont {K.~S.}\ \bibnamefont
  {Dawson}} \emph {et~al.} (\bibinfo {collaboration} {BOSS}),\ }\href
  {https://doi.org/10.1088/0004-6256/145/1/10} {\bibfield  {journal} {\bibinfo
  {journal} {Astron. J.}\ }\textbf {\bibinfo {volume} {145}},\ \bibinfo {pages}
  {10} (\bibinfo {year} {2013})},\ \Eprint {https://arxiv.org/abs/1208.0022}
  {arXiv:1208.0022 [astro-ph.CO]} \BibitemShut {NoStop}%
\bibitem [{\citenamefont {{Behroozi}}\ \emph {et~al.}(2013)\citenamefont
  {{Behroozi}}, \citenamefont {{Wechsler}},\ and\ \citenamefont
  {{Wu}}}]{2013ApJ...762..109B}%
  \BibitemOpen
  \bibfield  {author} {\bibinfo {author} {\bibfnamefont {P.~S.}\ \bibnamefont
  {{Behroozi}}}, \bibinfo {author} {\bibfnamefont {R.~H.}\ \bibnamefont
  {{Wechsler}}},\ and\ \bibinfo {author} {\bibfnamefont {H.-Y.}\ \bibnamefont
  {{Wu}}},\ }\href {https://doi.org/10.1088/0004-637X/762/2/109} {\bibfield
  {journal} {\bibinfo  {journal} {\apj}\ }\textbf {\bibinfo {volume} {762}},\
  \bibinfo {eid} {109} (\bibinfo {year} {2013})},\ \Eprint
  {https://arxiv.org/abs/1110.4372} {arXiv:1110.4372 [astro-ph.CO]}
  \BibitemShut {NoStop}%
\bibitem [{\citenamefont {Miyatake}\ \emph {et~al.}(2015)\citenamefont
  {Miyatake}, \citenamefont {More}, \citenamefont {Mandelbaum}, \citenamefont
  {Takada}, \citenamefont {Spergel}, \citenamefont {Kneib}, \citenamefont
  {Schneider}, \citenamefont {Brinkmann},\ and\ \citenamefont
  {Brownstein}}]{Miyatake:2013bha}%
  \BibitemOpen
  \bibfield  {author} {\bibinfo {author} {\bibfnamefont {H.}~\bibnamefont
  {Miyatake}}, \bibinfo {author} {\bibfnamefont {S.}~\bibnamefont {More}},
  \bibinfo {author} {\bibfnamefont {R.}~\bibnamefont {Mandelbaum}}, \bibinfo
  {author} {\bibfnamefont {M.}~\bibnamefont {Takada}}, \bibinfo {author}
  {\bibfnamefont {D.~N.}\ \bibnamefont {Spergel}}, \bibinfo {author}
  {\bibfnamefont {J.-P.}\ \bibnamefont {Kneib}}, \bibinfo {author}
  {\bibfnamefont {D.~P.}\ \bibnamefont {Schneider}}, \bibinfo {author}
  {\bibfnamefont {J.}~\bibnamefont {Brinkmann}},\ and\ \bibinfo {author}
  {\bibfnamefont {J.~R.}\ \bibnamefont {Brownstein}},\ }\href
  {https://doi.org/10.1088/0004-637X/806/1/1} {\bibfield  {journal} {\bibinfo
  {journal} {Astrophys. J.}\ }\textbf {\bibinfo {volume} {806}},\ \bibinfo
  {pages} {1} (\bibinfo {year} {2015})},\ \Eprint
  {https://arxiv.org/abs/1311.1480} {arXiv:1311.1480 [astro-ph.CO]}
  \BibitemShut {NoStop}%
\bibitem [{\citenamefont {{Tinker}}\ \emph {et~al.}(2010)\citenamefont
  {{Tinker}}, \citenamefont {{Robertson}}, \citenamefont {{Kravtsov}},
  \citenamefont {{Klypin}}, \citenamefont {{Warren}}, \citenamefont {{Yepes}},\
  and\ \citenamefont {{Gottl{\"o}ber}}}]{2010ApJ...724..878T}%
  \BibitemOpen
  \bibfield  {author} {\bibinfo {author} {\bibfnamefont {J.~L.}\ \bibnamefont
  {{Tinker}}}, \bibinfo {author} {\bibfnamefont {B.~E.}\ \bibnamefont
  {{Robertson}}}, \bibinfo {author} {\bibfnamefont {A.~V.}\ \bibnamefont
  {{Kravtsov}}}, \bibinfo {author} {\bibfnamefont {A.}~\bibnamefont
  {{Klypin}}}, \bibinfo {author} {\bibfnamefont {M.~S.}\ \bibnamefont
  {{Warren}}}, \bibinfo {author} {\bibfnamefont {G.}~\bibnamefont {{Yepes}}},\
  and\ \bibinfo {author} {\bibfnamefont {S.}~\bibnamefont {{Gottl{\"o}ber}}},\
  }\href {https://doi.org/10.1088/0004-637X/724/2/878} {\bibfield  {journal}
  {\bibinfo  {journal} {\apj}\ }\textbf {\bibinfo {volume} {724}},\ \bibinfo
  {pages} {878} (\bibinfo {year} {2010})},\ \Eprint
  {https://arxiv.org/abs/1001.3162} {arXiv:1001.3162 [astro-ph.CO]}
  \BibitemShut {NoStop}%
\bibitem [{\citenamefont {{Varshalovich}}\ \emph {et~al.}(1988)\citenamefont
  {{Varshalovich}}, \citenamefont {{Moskalev}},\ and\ \citenamefont
  {{Khersonskii}}}]{1988qtam.book.....V}%
  \BibitemOpen
  \bibfield  {author} {\bibinfo {author} {\bibfnamefont {D.~A.}\ \bibnamefont
  {{Varshalovich}}}, \bibinfo {author} {\bibfnamefont {A.~N.}\ \bibnamefont
  {{Moskalev}}},\ and\ \bibinfo {author} {\bibfnamefont {V.~K.}\ \bibnamefont
  {{Khersonskii}}},\ }\href {https://doi.org/10.1142/0270} {\emph {\bibinfo
  {title} {{Quantum Theory of Angular Momentum}}}}\ (\bibinfo {year}
  {1988})\BibitemShut {NoStop}%
\bibitem [{\citenamefont {Hartlap}\ \emph {et~al.}(2006)\citenamefont
  {Hartlap}, \citenamefont {Simon},\ and\ \citenamefont
  {Schneider}}]{Hartlap:2006kj}%
  \BibitemOpen
  \bibfield  {author} {\bibinfo {author} {\bibfnamefont {J.}~\bibnamefont
  {Hartlap}}, \bibinfo {author} {\bibfnamefont {P.}~\bibnamefont {Simon}},\
  and\ \bibinfo {author} {\bibfnamefont {P.}~\bibnamefont {Schneider}},\
  }\bibfield  {journal} {\bibinfo  {journal} {Astron. Astrophys.}\ }\href
  {https://doi.org/10.1051/0004-6361:20066170} {10.1051/0004-6361:20066170}
  (\bibinfo {year} {2006}),\ \bibinfo {note} {[Astron.
  Astrophys.464,399(2007)]},\ \Eprint {https://arxiv.org/abs/astro-ph/0608064}
  {arXiv:astro-ph/0608064 [astro-ph]} \BibitemShut {NoStop}%
\bibitem [{\citenamefont {Scoccimarro}\ \emph {et~al.}(1999)\citenamefont
  {Scoccimarro}, \citenamefont {Couchman},\ and\ \citenamefont
  {Frieman}}]{Scoccimarro:1999ed}%
  \BibitemOpen
  \bibfield  {author} {\bibinfo {author} {\bibfnamefont {R.}~\bibnamefont
  {Scoccimarro}}, \bibinfo {author} {\bibfnamefont {H.~M.~P.}\ \bibnamefont
  {Couchman}},\ and\ \bibinfo {author} {\bibfnamefont {J.~A.}\ \bibnamefont
  {Frieman}},\ }\href {https://doi.org/10.1086/307220} {\bibfield  {journal}
  {\bibinfo  {journal} {Astrophys. J.}\ }\textbf {\bibinfo {volume} {517}},\
  \bibinfo {pages} {531} (\bibinfo {year} {1999})},\ \Eprint
  {https://arxiv.org/abs/astro-ph/9808305} {arXiv:astro-ph/9808305 [astro-ph]}
  \BibitemShut {NoStop}%
\bibitem [{\citenamefont {{McDonald}}\ and\ \citenamefont
  {{Roy}}(2009)}]{2009JCAP...08..020M}%
  \BibitemOpen
  \bibfield  {author} {\bibinfo {author} {\bibfnamefont {P.}~\bibnamefont
  {{McDonald}}}\ and\ \bibinfo {author} {\bibfnamefont {A.}~\bibnamefont
  {{Roy}}},\ }\href {https://doi.org/10.1088/1475-7516/2009/08/020} {\bibfield
  {journal} {\bibinfo  {journal} {\jcap}\ }\textbf {\bibinfo {volume} {2009}},\
  \bibinfo {eid} {020} (\bibinfo {year} {2009})},\ \Eprint
  {https://arxiv.org/abs/0902.0991} {arXiv:0902.0991 [astro-ph.CO]}
  \BibitemShut {NoStop}%
\bibitem [{\citenamefont {{Bernardeau}}\ \emph {et~al.}(2002)\citenamefont
  {{Bernardeau}}, \citenamefont {{Colombi}}, \citenamefont {{Gazta{\~n}aga}},\
  and\ \citenamefont {{Scoccimarro}}}]{2002PhR...367....1B}%
  \BibitemOpen
  \bibfield  {author} {\bibinfo {author} {\bibfnamefont {F.}~\bibnamefont
  {{Bernardeau}}}, \bibinfo {author} {\bibfnamefont {S.}~\bibnamefont
  {{Colombi}}}, \bibinfo {author} {\bibfnamefont {E.}~\bibnamefont
  {{Gazta{\~n}aga}}},\ and\ \bibinfo {author} {\bibfnamefont {R.}~\bibnamefont
  {{Scoccimarro}}},\ }\href {https://doi.org/10.1016/S0370-1573(02)00135-7}
  {\bibfield  {journal} {\bibinfo  {journal} {\physrep}\ }\textbf {\bibinfo
  {volume} {367}},\ \bibinfo {pages} {1} (\bibinfo {year} {2002})},\ \Eprint
  {https://arxiv.org/abs/astro-ph/0112551} {arXiv:astro-ph/0112551 [astro-ph]}
  \BibitemShut {NoStop}%
\bibitem [{\citenamefont {Kaiser}(1987)}]{Kaiser:1987qv}%
  \BibitemOpen
  \bibfield  {author} {\bibinfo {author} {\bibfnamefont {N.}~\bibnamefont
  {Kaiser}},\ }\href@noop {} {\bibfield  {journal} {\bibinfo  {journal} {Mon.
  Not. Roy. Astron. Soc.}\ }\textbf {\bibinfo {volume} {227}},\ \bibinfo
  {pages} {1} (\bibinfo {year} {1987})}\BibitemShut {NoStop}%
\bibitem [{\citenamefont {Schmittfull}\ \emph {et~al.}(2015)\citenamefont
  {Schmittfull}, \citenamefont {Feng}, \citenamefont {Beutler}, \citenamefont
  {Sherwin},\ and\ \citenamefont {Chu}}]{Schmittfull:2015mja}%
  \BibitemOpen
  \bibfield  {author} {\bibinfo {author} {\bibfnamefont {M.}~\bibnamefont
  {Schmittfull}}, \bibinfo {author} {\bibfnamefont {Y.}~\bibnamefont {Feng}},
  \bibinfo {author} {\bibfnamefont {F.}~\bibnamefont {Beutler}}, \bibinfo
  {author} {\bibfnamefont {B.}~\bibnamefont {Sherwin}},\ and\ \bibinfo {author}
  {\bibfnamefont {M.~Y.}\ \bibnamefont {Chu}},\ }\href
  {https://doi.org/10.1103/PhysRevD.92.123522} {\bibfield  {journal} {\bibinfo
  {journal} {Phys. Rev.}\ }\textbf {\bibinfo {volume} {D92}},\ \bibinfo {pages}
  {123522} (\bibinfo {year} {2015})},\ \Eprint
  {https://arxiv.org/abs/1508.06972} {arXiv:1508.06972 [astro-ph.CO]}
  \BibitemShut {NoStop}%
\bibitem [{\citenamefont {Hikage}\ \emph {et~al.}(2017)\citenamefont {Hikage},
  \citenamefont {Koyama},\ and\ \citenamefont {Heavens}}]{Hikage:2017tmm}%
  \BibitemOpen
  \bibfield  {author} {\bibinfo {author} {\bibfnamefont {C.}~\bibnamefont
  {Hikage}}, \bibinfo {author} {\bibfnamefont {K.}~\bibnamefont {Koyama}},\
  and\ \bibinfo {author} {\bibfnamefont {A.}~\bibnamefont {Heavens}},\ }\href
  {https://doi.org/10.1103/PhysRevD.96.043513} {\bibfield  {journal} {\bibinfo
  {journal} {Phys. Rev.}\ }\textbf {\bibinfo {volume} {D96}},\ \bibinfo {pages}
  {043513} (\bibinfo {year} {2017})},\ \Eprint
  {https://arxiv.org/abs/1703.07878} {arXiv:1703.07878 [astro-ph.CO]}
  \BibitemShut {NoStop}%
\bibitem [{\citenamefont {Jeong}\ and\ \citenamefont
  {Komatsu}(2009)}]{Jeong:2009vd}%
  \BibitemOpen
  \bibfield  {author} {\bibinfo {author} {\bibfnamefont {D.}~\bibnamefont
  {Jeong}}\ and\ \bibinfo {author} {\bibfnamefont {E.}~\bibnamefont
  {Komatsu}},\ }\href {https://doi.org/10.1088/0004-637X/703/2/1230} {\bibfield
   {journal} {\bibinfo  {journal} {Astrophys. J.}\ }\textbf {\bibinfo {volume}
  {703}},\ \bibinfo {pages} {1230} (\bibinfo {year} {2009})},\ \Eprint
  {https://arxiv.org/abs/0904.0497} {arXiv:0904.0497 [astro-ph.CO]}
  \BibitemShut {NoStop}%
\bibitem [{\citenamefont {Sefusatti}(2009)}]{Sefusatti:2009qh}%
  \BibitemOpen
  \bibfield  {author} {\bibinfo {author} {\bibfnamefont {E.}~\bibnamefont
  {Sefusatti}},\ }\href {https://doi.org/10.1103/PhysRevD.80.123002} {\bibfield
   {journal} {\bibinfo  {journal} {Phys. Rev.}\ }\textbf {\bibinfo {volume}
  {D80}},\ \bibinfo {pages} {123002} (\bibinfo {year} {2009})},\ \Eprint
  {https://arxiv.org/abs/0905.0717} {arXiv:0905.0717 [astro-ph.CO]}
  \BibitemShut {NoStop}%
\bibitem [{\citenamefont {McDonald}(2006)}]{McDonald:2006mx}%
  \BibitemOpen
  \bibfield  {author} {\bibinfo {author} {\bibfnamefont {P.}~\bibnamefont
  {McDonald}},\ }\href {https://doi.org/10.1103/PhysRevD.74.103512,
  10.1103/PhysRevD.74.129901} {\bibfield  {journal} {\bibinfo  {journal} {Phys.
  Rev.}\ }\textbf {\bibinfo {volume} {D74}},\ \bibinfo {pages} {103512}
  (\bibinfo {year} {2006})},\ \bibinfo {note} {[Erratum: Phys.
  Rev.D74,129901(2006)]},\ \Eprint {https://arxiv.org/abs/astro-ph/0609413}
  {arXiv:astro-ph/0609413 [astro-ph]} \BibitemShut {NoStop}%
\bibitem [{\citenamefont {{Saito}}\ \emph {et~al.}(2014)\citenamefont
  {{Saito}}, \citenamefont {{Baldauf}}, \citenamefont {{Vlah}}, \citenamefont
  {{Seljak}}, \citenamefont {{Okumura}},\ and\ \citenamefont
  {{McDonald}}}]{Saito:2014aa}%
  \BibitemOpen
  \bibfield  {author} {\bibinfo {author} {\bibfnamefont {S.}~\bibnamefont
  {{Saito}}}, \bibinfo {author} {\bibfnamefont {T.}~\bibnamefont {{Baldauf}}},
  \bibinfo {author} {\bibfnamefont {Z.}~\bibnamefont {{Vlah}}}, \bibinfo
  {author} {\bibfnamefont {U.}~\bibnamefont {{Seljak}}}, \bibinfo {author}
  {\bibfnamefont {T.}~\bibnamefont {{Okumura}}},\ and\ \bibinfo {author}
  {\bibfnamefont {P.}~\bibnamefont {{McDonald}}},\ }\href
  {https://doi.org/10.1103/PhysRevD.90.123522} {\bibfield  {journal} {\bibinfo
  {journal} {\prd}\ }\textbf {\bibinfo {volume} {90}},\ \bibinfo {eid} {123522}
  (\bibinfo {year} {2014})},\ \Eprint {https://arxiv.org/abs/1405.1447}
  {arXiv:1405.1447 [astro-ph.CO]} \BibitemShut {NoStop}%
\bibitem [{\citenamefont {Desjacques}\ \emph {et~al.}(2018)\citenamefont
  {Desjacques}, \citenamefont {Jeong},\ and\ \citenamefont
  {Schmidt}}]{Desjacques:2016bnm}%
  \BibitemOpen
  \bibfield  {author} {\bibinfo {author} {\bibfnamefont {V.}~\bibnamefont
  {Desjacques}}, \bibinfo {author} {\bibfnamefont {D.}~\bibnamefont {Jeong}},\
  and\ \bibinfo {author} {\bibfnamefont {F.}~\bibnamefont {Schmidt}},\ }\href
  {https://doi.org/10.1016/j.physrep.2017.12.002} {\bibfield  {journal}
  {\bibinfo  {journal} {Phys. Rept.}\ }\textbf {\bibinfo {volume} {733}},\
  \bibinfo {pages} {1} (\bibinfo {year} {2018})},\ \Eprint
  {https://arxiv.org/abs/1611.09787} {arXiv:1611.09787 [astro-ph.CO]}
  \BibitemShut {NoStop}%
\bibitem [{\citenamefont {{Moradinezhad Dizgah}}\ \emph
  {et~al.}(2020)\citenamefont {{Moradinezhad Dizgah}}, \citenamefont
  {{Biagetti}}, \citenamefont {{Sefusatti}}, \citenamefont {{Desjacques}},\
  and\ \citenamefont {{Nore{\~n}a}}}]{2020arXiv201014523M}%
  \BibitemOpen
  \bibfield  {author} {\bibinfo {author} {\bibfnamefont {A.}~\bibnamefont
  {{Moradinezhad Dizgah}}}, \bibinfo {author} {\bibfnamefont {M.}~\bibnamefont
  {{Biagetti}}}, \bibinfo {author} {\bibfnamefont {E.}~\bibnamefont
  {{Sefusatti}}}, \bibinfo {author} {\bibfnamefont {V.}~\bibnamefont
  {{Desjacques}}},\ and\ \bibinfo {author} {\bibfnamefont {J.}~\bibnamefont
  {{Nore{\~n}a}}},\ }\href@noop {} {\bibfield  {journal} {\bibinfo  {journal}
  {arXiv e-prints}\ ,\ \bibinfo {eid} {arXiv:2010.14523}} (\bibinfo {year}
  {2020})},\ \Eprint {https://arxiv.org/abs/2010.14523} {arXiv:2010.14523
  [astro-ph.CO]} \BibitemShut {NoStop}%
\bibitem [{\citenamefont {{Lazeyras}}\ \emph {et~al.}(2016)\citenamefont
  {{Lazeyras}}, \citenamefont {{Wagner}}, \citenamefont {{Baldauf}},\ and\
  \citenamefont {{Schmidt}}}]{2016JCAP...02..018L}%
  \BibitemOpen
  \bibfield  {author} {\bibinfo {author} {\bibfnamefont {T.}~\bibnamefont
  {{Lazeyras}}}, \bibinfo {author} {\bibfnamefont {C.}~\bibnamefont
  {{Wagner}}}, \bibinfo {author} {\bibfnamefont {T.}~\bibnamefont
  {{Baldauf}}},\ and\ \bibinfo {author} {\bibfnamefont {F.}~\bibnamefont
  {{Schmidt}}},\ }\href {https://doi.org/10.1088/1475-7516/2016/02/018}
  {\bibfield  {journal} {\bibinfo  {journal} {\jcap}\ }\textbf {\bibinfo
  {volume} {2016}},\ \bibinfo {eid} {018} (\bibinfo {year} {2016})},\ \Eprint
  {https://arxiv.org/abs/1511.01096} {arXiv:1511.01096 [astro-ph.CO]}
  \BibitemShut {NoStop}%
\bibitem [{\citenamefont {{Sheth}}\ \emph {et~al.}(2013)\citenamefont
  {{Sheth}}, \citenamefont {{Chan}},\ and\ \citenamefont
  {{Scoccimarro}}}]{2013PhRvD..87h3002S}%
  \BibitemOpen
  \bibfield  {author} {\bibinfo {author} {\bibfnamefont {R.~K.}\ \bibnamefont
  {{Sheth}}}, \bibinfo {author} {\bibfnamefont {K.~C.}\ \bibnamefont
  {{Chan}}},\ and\ \bibinfo {author} {\bibfnamefont {R.}~\bibnamefont
  {{Scoccimarro}}},\ }\href {https://doi.org/10.1103/PhysRevD.87.083002}
  {\bibfield  {journal} {\bibinfo  {journal} {\prd}\ }\textbf {\bibinfo
  {volume} {87}},\ \bibinfo {eid} {083002} (\bibinfo {year} {2013})},\ \Eprint
  {https://arxiv.org/abs/1207.7117} {arXiv:1207.7117 [astro-ph.CO]}
  \BibitemShut {NoStop}%
\bibitem [{\citenamefont {{Castorina}}\ \emph {et~al.}(2019)\citenamefont
  {{Castorina}}, \citenamefont {{Hand}}, \citenamefont {{Seljak}},
  \citenamefont {{Beutler}}, \citenamefont {{Chuang}}, \citenamefont {{Zhao}},
  \citenamefont {{Gil-Mar{\'\i}n}}, \citenamefont {{Percival}}, \citenamefont
  {{Ross}}, \citenamefont {{Choi}}, \citenamefont {{Dawson}}, \citenamefont
  {{de la Macorra}}, \citenamefont {{Rossi}}, \citenamefont {{Ruggeri}},
  \citenamefont {{Schneider}},\ and\ \citenamefont
  {{Zhao}}}]{2019JCAP...09..010C}%
  \BibitemOpen
  \bibfield  {author} {\bibinfo {author} {\bibfnamefont {E.}~\bibnamefont
  {{Castorina}}}, \bibinfo {author} {\bibfnamefont {N.}~\bibnamefont {{Hand}}},
  \bibinfo {author} {\bibfnamefont {U.}~\bibnamefont {{Seljak}}}, \bibinfo
  {author} {\bibfnamefont {F.}~\bibnamefont {{Beutler}}}, \bibinfo {author}
  {\bibfnamefont {C.-H.}\ \bibnamefont {{Chuang}}}, \bibinfo {author}
  {\bibfnamefont {C.}~\bibnamefont {{Zhao}}}, \bibinfo {author} {\bibfnamefont
  {H.}~\bibnamefont {{Gil-Mar{\'\i}n}}}, \bibinfo {author} {\bibfnamefont
  {W.~J.}\ \bibnamefont {{Percival}}}, \bibinfo {author} {\bibfnamefont
  {A.~J.}\ \bibnamefont {{Ross}}}, \bibinfo {author} {\bibfnamefont {P.~D.}\
  \bibnamefont {{Choi}}}, \bibinfo {author} {\bibfnamefont {K.}~\bibnamefont
  {{Dawson}}}, \bibinfo {author} {\bibfnamefont {A.}~\bibnamefont {{de la
  Macorra}}}, \bibinfo {author} {\bibfnamefont {G.}~\bibnamefont {{Rossi}}},
  \bibinfo {author} {\bibfnamefont {R.}~\bibnamefont {{Ruggeri}}}, \bibinfo
  {author} {\bibfnamefont {D.}~\bibnamefont {{Schneider}}},\ and\ \bibinfo
  {author} {\bibfnamefont {G.-B.}\ \bibnamefont {{Zhao}}},\ }\href
  {https://doi.org/10.1088/1475-7516/2019/09/010} {\bibfield  {journal}
  {\bibinfo  {journal} {\jcap}\ }\textbf {\bibinfo {volume} {2019}},\ \bibinfo
  {eid} {010} (\bibinfo {year} {2019})},\ \Eprint
  {https://arxiv.org/abs/1904.08859} {arXiv:1904.08859 [astro-ph.CO]}
  \BibitemShut {NoStop}%
\bibitem [{\citenamefont {{Hill}}(2018)}]{2018PhRvD..98h3542H}%
  \BibitemOpen
  \bibfield  {author} {\bibinfo {author} {\bibfnamefont {J.~C.}\ \bibnamefont
  {{Hill}}},\ }\href {https://doi.org/10.1103/PhysRevD.98.083542} {\bibfield
  {journal} {\bibinfo  {journal} {\prd}\ }\textbf {\bibinfo {volume} {98}},\
  \bibinfo {eid} {083542} (\bibinfo {year} {2018})},\ \Eprint
  {https://arxiv.org/abs/1807.07324} {arXiv:1807.07324 [astro-ph.CO]}
  \BibitemShut {NoStop}%
\end{thebibliography}%

\end{document}